\pdfoutput=1

\documentclass[11pt,twoside,a4paper,cmspaper,final,collab]{cms-tdr}
\def\svnVersion{316224}\def\cmsCernNo{2013-037}\def\cmsCernDate{\today}\def\cmsMessage{Published in Physics Letters B as \href{http://dx.doi.org/10.1016/j.physletb.2015.12.020}{\doi{10.1016/j.physletb.2015.12.020.}}}

\begin{document}\cmsNoteHeader{BPH-13-010}

\hyphenation{had-ron-i-za-tion}
\hyphenation{cal-or-i-me-ter}
\hyphenation{de-vices}
\RCS$Revision: 312025 $
\RCS$HeadURL: svn+ssh://svn.cern.ch/reps/tdr2/papers/BPH-13-010/trunk/BPH-13-010.tex $
\RCS$Id: BPH-13-010.tex 312025 2015-11-27 06:08:09Z stenson $
\newlength\cmsFigWidth
\ifthenelse{\boolean{cms@external}}{\setlength\cmsFigWidth{0.85\columnwidth}}{\setlength\cmsFigWidth{0.4\textwidth}}
\ifthenelse{\boolean{cms@external}}{\providecommand{\cmsLeft}{top\xspace}}{\providecommand{\cmsLeft}{left\xspace}}
\ifthenelse{\boolean{cms@external}}{\providecommand{\cmsRight}{bottom\xspace}}{\providecommand{\cmsRight}{right\xspace}}
\ifthenelse{\boolean{cms@external}}{\providecommand{\cmsBottom}{bottom}}{\providecommand{\cmsBottom}{bottom}}
\ifthenelse{\boolean{cms@external}}{\providecommand{\cmsLeft}{top}}{\providecommand{\cmsLeft}{left}}
\ifthenelse{\boolean{cms@external}}{\providecommand{\cmsRight}{bottom}}{\providecommand{\cmsRight}{right}}
\providecommand{\cPKst}{\ensuremath{\cmsSymbolFace{K}^\ast}\xspace}
\providecommand{\cPKstz}{\ensuremath{\cmsSymbolFace{K}^{\ast0}}\xspace}
\providecommand{\cPAKstz}{\ensuremath{\overline{\cmsSymbolFace{K}}{}^{\ast0}}\xspace}
\newcommand{\BtoKstmumu}{\ensuremath{\PBz\to\cPKstz \Pgmp \Pgmm}\xspace}
\newcommand{\BtoKstJpsi}{\ensuremath{\PBz\to\cPJgy \cPKstz}\xspace}
\newcommand{\BtoKstpsip}{\ensuremath{\PBz\to\psi' \cPKstz}\xspace}
\newcommand{\BtoKstJpsimumu}{\ensuremath{\PBz\to\cPJgy(\Pgmp \Pgmm \cPKstz})\xspace}
\newcommand{\BtoKstpsipmumu}{\ensuremath{\PBz\to\psi'(\Pgmp \Pgmm) \cPKstz}\xspace}
\newcommand{\BtoKstmumudecay}{\ensuremath{\PBz\to\cPKstz(\PKp \Pgpm) \Pgmp \Pgmm}\xspace}
\newcommand{\BtoKstJpsidecay}{\ensuremath{\PBz\to\cPJgy(\Pgmp \Pgmm) \cPKstz(\PKp \Pgpm)}\xspace}
\newcommand{\BtoKstpsipdecay}{\ensuremath{\PBz\to\psi'(\Pgmp \Pgmm) \cPKstz(\PKp \Pgpm)}\xspace}
\newcommand{\Kstmumudecay}{\ensuremath{\cPKstz(\PKp \Pgpm) \Pgmp \Pgmm}\xspace}
\newcommand{\KstJpsidecay}{\ensuremath{\cPJgy(\Pgmp \Pgmm) \cPKstz(\PKp \Pgpm)}\xspace}
\newcommand{\Kstpsipdecay}{\ensuremath{\psi'(\Pgmp \Pgmm) \cPKstz(\PKp \Pgpm)}\xspace}
\renewcommand{\PaBz}{\ensuremath{\overline{\cmsSymbolFace{B}}{}^{0}}\xspace}
\cmsNoteHeader{BPH-13-010}
\title{Angular analysis of the decay $\PBz \to \cPKstz \Pgmp \Pgmm$ from pp collisions at $\sqrt{s}=8\TeV$}

\date{\today}

\abstract{The angular distributions and the differential branching fraction of the decay $\PBz \to
  \PKst{}^0 \Pgmp \Pgmm$ are studied using data corresponding to an integrated luminosity
  of 20.5\fbinv collected with the CMS detector at the LHC in pp collisions at $\sqrt{s} = 8$\TeV.
  From 1430 signal decays, the forward-backward asymmetry of the muons, the $\PKst{}^0$
  longitudinal polarization fraction, and the differential branching fraction are determined as a
  function of the dimuon invariant mass squared. The measurements are among the most precise to date and are in good agreement with
  standard model predictions.}

\hypersetup{%
pdfauthor={CMS Collaboration},%
pdftitle={Angular analysis of the decay B0 to K*0 mu mu from pp collisions at sqrt(s) = 8 TeV},%
pdfsubject={CMS},%
pdfkeywords={CMS, physics, B0 decays}}

\maketitle
\section{Introduction}
\label{sec:Intro}

Phenomena beyond the standard model (SM) of particle physics can manifest themselves directly, via
the production of new particles, or indirectly, by affecting the production and decay of SM particles.  Analyses of
flavor-changing neutral current (FCNC) decays are particularly sensitive to the effect of new
physics, since such decays are highly suppressed in the SM\@.
The FCNC decay, \BtoKstmumu ($\cPKstz$ indicates the $\PKst{}^0$, and charge-conjugate states are
implied for all particles unless stated otherwise), provides many opportunities to search for new
phenomena.  In addition to the branching fraction, other properties of the decay can be measured,
including the forward-backward asymmetry of the muons, $A_\mathrm{FB}$, and the longitudinal
polarization fraction of the $\cPKstz$, $F_\mathrm{L}$.  To better understand this decay,
these quantities can be measured as a function of the dimuon invariant mass squared $(q^2)$.  New
physics may modify any of these quantities~\cite{Altmannshofer:2008dz, Melikhov:1998cd, Ali:1999mm,
  Yan:2000dc, Buchalla:2000sk, Feldmann:2002iw, Hiller:2003js, Kruger:2005ep, Hovhannisyan:2007pb,
  Egede:2008uy, Hurth:2008jc, Alok:2009tz, Alok:2010zd, Chang:2010zy, DescotesGenon:2011yn,
  Matias:2012xw,DescotesGenon:2012zf} relative to their SM values~\cite{Bobeth:2010wg,
  Bobeth:2011nj, Bobeth:2012vn, Ali:2006ew, Altmannshofer:2008dz, Altmannshofer:2011gn,
  Jager:2012uw, Descotes-Genon:2013vna}.  While previous measurements by BaBar, Belle, CDF, LHCb,
and CMS are consistent with the SM~\cite{BaBar, Belle, CDF, LHCb, Chatrchyan:2013cda}, they are
still statistically limited, and more precise measurements offer the possibility to uncover physics
beyond the SM\@.

In this Letter, we present measurements of $A_\mathrm{FB}$, $F_\mathrm{L}$, and the differential
branching fraction $\rd{}\mathcal{B}/\rd{}q^2$ from \BtoKstmumu decays, using data collected from pp
collisions at the CERN LHC by the CMS experiment at a center-of-mass energy of 8\TeV.  The
data correspond to an integrated luminosity of $20.5\pm0.5\fbinv$~\cite{LUMI}.  The
$\cPKstz$ is reconstructed through its decay to $\PKp\Pgpm$, and the $\PBz$ is reconstructed by
fitting the two identified muon tracks and the two hadron tracks to a common vertex. The values of
$A_\mathrm{FB}$ and $F_\mathrm{L}$ are measured by fitting the distribution of events as a function
of two angular variables: the angle between the positively charged muon and the $\PBz$ in the dimuon
rest frame, and the angle between the $\PKp$ and the $\PBz$ in the $\cPKstz$ rest frame.  All
measurements are performed in $q^2$ bins from 1 to $19\GeV^2$.  The $q^2$ bins
$8.68<q^2<10.09\GeV^2$ and $12.90<q^2<14.18\GeV^2$, corresponding to the \BtoKstJpsi and \BtoKstpsip
decays ($\psi'$ refers to the \Pgy), respectively, are used to validate the analysis.  The
former is also used to normalize the differential branching fraction.

\section{CMS detector}
\label{sec:Detector}

A detailed description of the CMS detector, together with a definition of the coordinate system used
and the standard kinematic variables, can be found in Ref.~\cite{CMS}.  The main detector components
used in this analysis are the silicon tracker and the muon detection systems. The silicon tracker,
located in the 3.8\unit{T} field of a superconducting solenoid, consists of three pixel layers and
ten strip layers (four of which have a stereo view) in the barrel region accompanied by similar
endcap pixel and strip detectors on each side that extend coverage out to $\abs{\eta}<2.5$.
For tracks with transverse momenta $1 < \pt < 10\GeV$ and $\abs{\eta} < 1.4$, the resolutions are
typically 1.5\% in \pt and 25--90 (45--150)\mum in the transverse (longitudinal) impact
parameter~\cite{TRK-11-001}.  Muons are measured in the range $\abs{\eta}< 2.4$, with
detection planes made using three technologies: drift tubes, cathode strip chambers, and
resistive plate chambers~\cite{Chatrchyan:2012xi}.  In addition to the tracker and muon detectors,
CMS is equipped with electromagnetic and hadronic calorimeters that cover $\abs{\eta}<5$.

Events are selected using a two-level trigger system. The first level has specialized hardware
processors that use information from the calorimeters and muon systems to select the most
interesting events. A high-level trigger processor farm further decreases the event rate from
around 90\unit{kHz} to around 400\unit{Hz}, before data storage.

\section{Reconstruction, event selection, and efficiency}
\label{sec:Selection}

The criteria used to select the candidate events during data taking (trigger) and after full event
reconstruction take advantage of the fact that $\PBz$ mesons have relatively long lifetimes and
therefore decay on average about 1\unit{mm} from their production point.  The trigger only uses muons to
select events, while the offline selection includes the full reconstruction of all decay products.

All events used in this analysis were recorded with the same trigger, requiring two identified muons
of opposite charge to form a vertex that is displaced from the pp collision region (beamspot).  The
beamspot position (most probable collision point) and size (the extent of the luminous region covering 68\% of the
collisions in each dimension) were continuously measured through Gaussian fits to reconstructed
vertices as part of the online data quality monitoring.  The trigger required each muon to have $\pt
> 3.5\GeV$, $\abs{\eta}<2.2$, and to pass within 2\,cm of the beam axis.  The dimuon system was
required to have $\pt > 6.9\GeV$, a vertex fit $\chi^2$ probability larger than 10\%, and a
separation of the vertex relative to the beamspot in the transverse plane of at least 3$\sigma$,
where $\sigma$ includes the calculated uncertainty in the vertex position and the measured size of
the beamspot.  In addition, the cosine of the angle, in the transverse plane, between the dimuon
momentum vector and the vector from the beamspot to the dimuon vertex was required to be greater
than 0.9.

The offline reconstruction requires two muons of opposite charge and two oppositely
charged hadrons.  The muons are required to match those that triggered the event readout, and
also to pass general muon identification requirements.  These include a track matched to at least one muon
segment (collection of hits in a muon chamber consistent with the passage of a charged particle), a
track fit $\chi^2$ per degree of freedom less than 1.8, hits in at least six tracker layers with at
least two from the pixel detector, and a transverse (longitudinal) impact parameter with respect to the beamspot less than 3\cm
(30\cm).  The reconstructed dimuon system must also satisfy the same requirements that
were applied in the trigger.

The hadron tracks are required to fail the muon identification criteria, have
$\pt>0.8\GeV$, and have an extrapolated distance of closest approach to the beamspot in
the transverse plane greater than twice the sum in quadrature of the distance uncertainty and the
beamspot transverse size.
The two hadrons must have an invariant mass within 90\MeV of the accepted $\cPKstz$ mass~\cite{PDG}
for either the $\PKp\Pgpm$ or $\PKm\Pgpp$ hypothesis.  To remove contamination from
$\Pgf\to\PKp\PKm$ decays, the invariant mass of the hadron pair must be greater than 1.035\GeV when
the charged kaon mass is assigned to both hadrons.  The $\PBz$ candidates are obtained by fitting
the four charged tracks to a common vertex, and applying a vertex constraint to improve the
resolution of the track parameters.  The $\PBz$ candidates must have $\pt>8\GeV$,
$\abs{\eta}<2.2$, vertex fit $\chi^2$ probability larger than 10\%, vertex transverse separation
from the beamspot greater than 12 times the sum in quadrature of the separation uncertainty and the
beamspot transverse size, and $\cos{\alpha_{xy}}>0.9994$, where $\alpha_{xy}$ is the angle, in the
transverse plane, between the $\PBz$ momentum vector and the line-of-flight between the beamspot and
the $\PBz$ vertex.  The invariant mass $m$ of the $\PBz$ candidate must also be within 280\MeV of
the accepted $\PBz$ mass $m_\PBz$~\cite{PDG} for either the $\PKm\Pgpp\Pgmp\Pgmm$ or
$\PKp\Pgpm\Pgmp\Pgmm$ hypothesis.  The selection criteria are optimized using simulated signal
samples (described below) and background from data using sidebands of the $\PBz$ mass.  After applying the selection criteria,
events in which at least one candidate is found contain on average 1.05 candidates.  A single
candidate is chosen from each event based on the best $\PBz$ vertex fit $\chi^2$.

From the selected events, the dimuon invariant mass $q$ and its calculated uncertainty $\sigma_{q}$
are used to distinguish the signal from the control samples.  The control samples \BtoKstJpsi and
\BtoKstpsip are defined by $\abs{q - m_{\cPJgy}} < 3\sigma_{q}$ and $\abs{q - m_{\psi'}} <
3\sigma_{q}$, respectively, where $m_{\cPJgy}$ and $m_{\psi'}$ are the accepted masses~\cite{PDG}.  The average
value for $\sigma_{q}$ is about 26\MeV.  The signal sample is composed of the events that are not
assigned to the $\cPJgy$ and $\psi'$ samples.

The signal sample still contains contributions from the control samples, mainly due to
unreconstructed soft photons in the charmonium decay.  These events will have a low $q$ value
and fall outside the selection described above.  These events will also have a low $m$ value and
therefore they can be selectively removed using a combined selection on $q$ and $m$.  For
$q<m_{\cPJgy}$ $(q>m_{\cPJgy})$, we require $\abs{(m-m_{B^0})-(q-m_{\cPJgy})}>160\: (60)\MeV$.  For
$q<m_{\psi'}$ $(q>m_{\psi'})$, we require $\abs{(m-m_{B^0})-(q-m_{\psi'})}>60\: (30)\MeV$.  The
requirements are set such that less than 10\% of the background events originate from the
control channels.

The four-track vertex candidate is identified as a $\PBz$ or $\PaBz$ depending on whether the $\PKp\Pgpm$
or $\PKm\Pgpp$ invariant mass is closest to the accepted $\cPKstz$ mass.  The fraction of
candidates assigned to the incorrect state is estimated from simulations to be 12--14\%, depending on
$q^2$.

The global efficiency, $\epsilon$, is the product of the acceptance and the combined trigger,
reconstruction, and selection efficiency, both of which are obtained from Monte Carlo (MC)
simulations.  The pp collisions are simulated using \PYTHIA~\cite{Pythia} version 6.424, the
unstable particles are decayed by \EVTGEN~\cite{EvtGen} version 9.1 (using the default matrix
element for the signal), and the particles are propagated through a detailed model of the detector with
\GEANTfour~\cite{Geant4}.  The reconstruction and selection of the generated events proceed as for
data.  Three simulated samples were created in which the $\PBz$ was forced to decay to
\Kstmumudecay, \KstJpsidecay, or \Kstpsipdecay.  The samples were constructed to
ensure that the number and spatial distribution of pp collision vertices in each event match the
distributions found in data.  The acceptance is obtained from generated events, before the particle
propagation with \GEANTfour, and is calculated as the fraction of events passing the single-muon
requirement of $\pt(\mu)>3.3\GeV$ and $\abs{\eta(\mu)}<2.3$ relative to all events with
$\pt(\PBz)>8\GeV$ and $\abs{\eta(\PBz)}<2.2$.  As the acceptance requirements are placed on the
generated quantities, they are less restrictive than the final selection requirements, which are
based on the reconstructed quantities, to allow for the effect of finite resolution.  Only events passing the
acceptance criteria are processed through the \GEANT simulation, the trigger simulation, and the
reconstruction software.  The combined trigger, reconstruction, and selection efficiency is the
ratio of the number of events that pass the trigger and selection requirements and have a
reconstructed $\PBz$ compatible with the generated $\PBz$ in the event, relative to the number of
events that pass the acceptance criteria.  The compatibility of generated and reconstructed
particles is enforced by requiring the reconstructed $\PKp$, $\Pgpm$, $\Pgmp$, and $\Pgmm$ to have
$\sqrt{(\Delta \eta)^2 + (\Delta \varphi)^2}$ less than 0.3 (0.004) for hadrons (muons), where
$\Delta \eta$ and $\Delta \varphi$ are the differences in $\eta$ and $\varphi$ between the
reconstructed and generated particles.  Requiring all four particles in the $\PBz$ decay to be
matched results in an efficiency of 99.6\% (0.4\% of the events have a correctly reconstructed
$\PBz$ that is not matched to a generated $\PBz$) and a purity of 99.5\% (0.5\% of the matched
candidates are not a correctly reconstructed $\PBz$).  Efficiencies are determined for both
correctly tagged (the $\PK$ and $\Pgp$ have the correct charge) and mistagged (the $\PK$ and
$\Pgp$ charges are reversed) candidates.

\section{Analysis method}
\label{sec:Analysis}

\begin{figure*}[htbp]
  \begin{center}
    \includegraphics[width=0.99\textwidth]{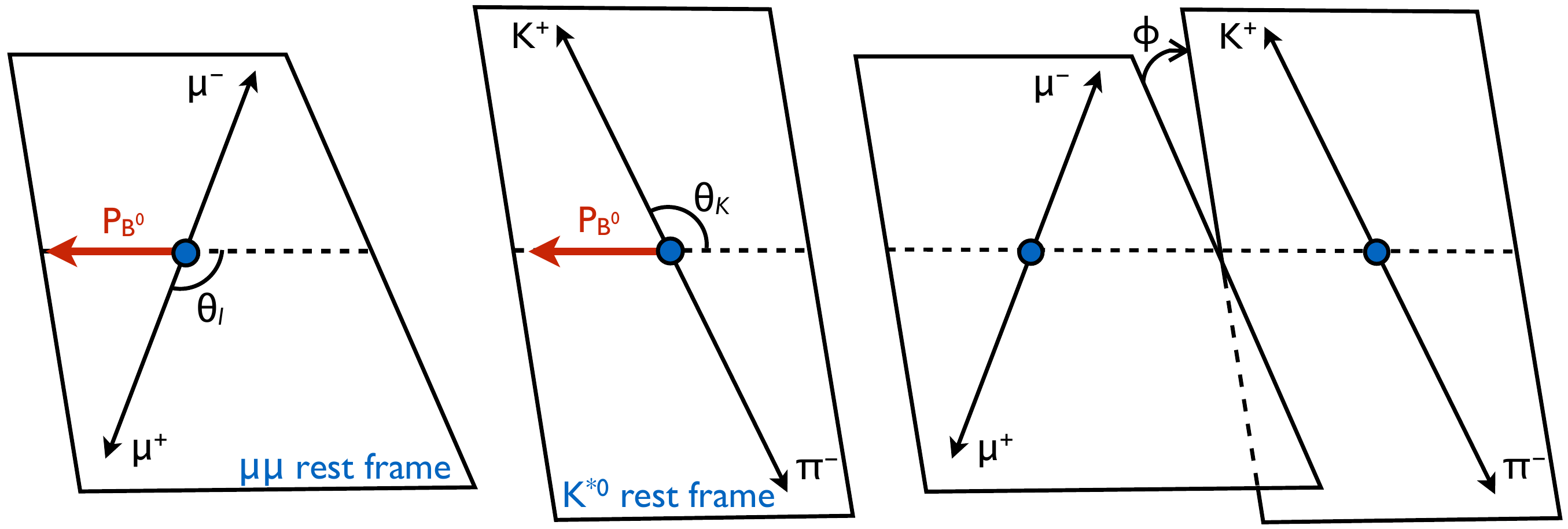}
    \caption{Sketch showing the definition of the angular observables $\theta_l$ (left), $\theta_\PK$ (middle), and $\phi$ (right) for the decay \BtoKstmumudecay.}
    \label{fig:ske}
  \end{center}
\end{figure*}

This analysis measures $A_\mathrm{FB}$, $F_\mathrm{L}$, and $\rd{}\mathcal{B}/\rd{}q^2$ of the decay
\BtoKstmumu as a function of $q^2$.  Figure~\ref{fig:ske} shows the angular observables needed to
define the decay: $\theta_\PK$ is the angle between the kaon momentum and the direction opposite to
the $\PBz$ $\big(\PaBz\big)$ in the $\cPKstz$ $\big(\cPAKstz\big)$ rest frame, $\theta_l$ is the
angle between the positive (negative) muon momentum and the direction opposite to the $\PBz$
$\big(\PaBz\big)$ in the dimuon rest frame, and $\phi$ is the angle between the plane containing the
two muons and the plane containing the kaon and pion.  As the extracted angular parameters
$A_\mathrm{FB}$ and $F_\mathrm{L}$ do not depend on $\phi$ and the product of the acceptance and
efficiency is nearly constant as a function of $\phi$, the angle $\phi$ is integrated out.  Although
the $\PKp\Pgpm$ invariant mass must be consistent with that of a $\cPKstz$, there can be a contribution from
spinless (S-wave) $\PKp\Pgpm$
combinations~\cite{Descotes-Genon:2013vna,Becirevic:2012dp,Matias:2012qz,Blake:Swave}.  This is
parametrized with two terms: $F_\mathrm{S}$, which is related to the S-wave fraction, and
$A_\mathrm{S}$, which is the interference amplitude between the S-wave and P-wave decays.  Including
this component, the angular distribution of \BtoKstmumu can be written
as~\cite{Descotes-Genon:2013vna}:
\ifthenelse{\boolean{cms@external}}{
\begin{multline}\label{eq:angALL}
  \frac{1}{\Gamma}\frac{\rd{}^3\Gamma}{\rd{}\cos\theta_\PK\, \rd{}\cos\theta_l\, \rd{}q^2}  = \\
  \begin{aligned}
   \qquad&\frac{9}{16} \left\lbrace \frac{2}{3} \Bigl[ F_\mathrm{S} +  A_\mathrm{S} \cos\theta_\PK \Bigr] \left(1 - \cos^2\theta_l\right) \right. \\
    & \left. +\; \left(1 - F_\mathrm{S}\right) \Bigl[2 F_\mathrm{L} \cos^2\theta_\PK \left(1 - \cos^2\theta_l\right) \right. \\
    & \left. +\; \frac{1}{2} \left(1 - F_\mathrm{L}\right) \left(1 - \cos^2\theta_\PK\right) \left(1 + \cos^2\theta_l\right) \right. \\
    & \left. +\; \frac{4}{3} A_\mathrm{FB} \left(1 - \cos^2\theta_\PK\right) \cos\theta_l\Bigr] \right\rbrace.
  \end{aligned}
\end{multline}
}{
\begin{equation}\label{eq:angALL}
  \begin{split}
     \frac{1}{\Gamma}\frac{\rd{}^3\Gamma}{\rd{}\!\cos\theta_\PK\, \rd{}\!\cos\theta_l\, \rd{}q^2}  &=
      \frac{9}{16} \left\lbrace \frac{2}{3} \Bigl[ F_\mathrm{S} + A_\mathrm{S} \cos\theta_\PK \Bigr] \left(1 - \cos^2\theta_l\right) \right. \\
    & \left. +\; \left(1 - F_\mathrm{S}\right) \Bigl[2 F_\mathrm{L} \cos^2\theta_\PK \left(1 - \cos^2\theta_l\right) \right. \\
    & \left. +\; \frac{1}{2} \left(1 - F_\mathrm{L}\right) \left(1 - \cos^2\theta_\PK\right) \left(1 + \cos^2\theta_l\right) \right. \\
    & \left. +\; \frac{4}{3} A_\mathrm{FB} \left(1 - \cos^2\theta_\PK\right) \cos\theta_l\Bigr] \right\rbrace.
   \end{split}
\end{equation}
}

For each $q^2$ bin, the observables of interest are extracted from an unbinned extended maximum-likelihood fit
to three variables: the $\PKp\Pgpm\Pgmp\Pgmm$ invariant mass $m$ and the two angular
variables ${\theta_\PK}$ and ${\theta_l}$.  For each $q^2$ bin, the unnormalized probability density function
(PDF) has the following expression:
\begin{equation} \label{eq:PDF}
  \begin{split}
    \mathrm{PDF}(m,\theta_\PK,\theta_l) & = Y^{C}_{S} \biggl[ S^{C}(m)  \, S^a(\theta_\PK,\theta_l) \, \epsilon^{C}(\theta_\PK,\theta_l) \biggr. \\
    & \biggl. + \frac{f^{M}}{1-f^{M}}~S^{M}(m) \, S^a(-\theta_\PK,-\theta_l) \, \epsilon^{M}(\theta_\PK,\theta_l) \biggr] \\
    & + Y_{B}\,B^m(m) \, B^{\theta_\PK}(\theta_\PK) \, B^{\theta_l}(\theta_l), \\
  \end{split}
\end{equation}
where the contributions correspond to correctly tagged signal events, mistagged signal events, and
background events.  The parameters $Y^{C}_{S}$ and $Y_{B}$ are the yields of correctly tagged signal
events and background events, respectively, and are free parameters in the fit.  The parameter
$f^{M}$ is the fraction of signal events that are mistagged and is determined from MC simulation.
The signal mass probability functions $S^{C}(m)$ and $S^{M}(m)$ are each the sum of two Gaussian functions and describe the
mass distribution for correctly tagged and mistagged signal events, respectively.  In the fit, there
is one free parameter for the mass value in both signal functions, while the other parameters (four
Gaussian $\sigma$ parameters and two fractions relating the contribution of each Gaussian) are obtained
from MC simulation, which has been found to accurately reproduce the data.
The function $S^a(\theta_\PK,\theta_l)$ describes the signal in the
two-dimensional (2D) space of the angular observables and corresponds to Eq.~(\ref{eq:angALL}).  The
combination $B^m(m) \, B^{\theta_\PK}(\theta_\PK) \, B^{\theta_l}(\theta_l)$ is obtained from
$\PBz$ sideband data and describes the background in the space of $(m,\theta_\PK,\theta_l)$,
where the mass distribution is an exponential function and the angular distributions are polynomials ranging from
second to fourth degree, depending on the $q^2$ bin and the angular variable.  The functions
$\epsilon^{C}(\theta_\PK,\theta_l)$ and $\epsilon^{M}(\theta_\PK,\theta_l)$ are
the efficiencies in the 2D space of $-1\leq \cos\theta_\PK\leq1,-1\leq \cos\theta_l\leq 1$ for
correctly tagged and mistagged signal
events, respectively.  The efficiency function for correctly tagged events is obtained from a fit to
the 2D-binned efficiency from simulation and is constrained to be positive.  There are 30 bins (5 in
$\cos\theta_\PK$ and 6 in $\cos\theta_l$), and the efficiency fit function is a polynomial of third degree in
$\cos\theta_\PK$ and fifth degree in $\cos\theta_l$ (and all cross terms), for a total of 24 free
parameters.  This procedure does not work for the mistagged events because of the much smaller number of
events (resulting in empty bins) and a more complicated efficiency.  For mistagged events, the
2D efficiency is calculated in $5{\times}5$ bins of $\cos\theta_\PK$ and $\cos\theta_l$, and an
interpolation is performed.
This interpolation function is used to generate a new binned efficiency (in $120{\times}120$
bins), with all bin contents constrained to be nonnegative.
The efficiency function uses this finely binned efficiency, with linear interpolation between bins.
The efficiencies for both correctly tagged and mistagged events peak at $\cos\theta_l$ near 0 for
$q^2<10\GeV^2$, becoming flat for larger values of $q^2$.  The efficiency for correctly tagged
events tends to decrease with increasing $\cos\theta_K$, and for $q^2>14\GeV^2$ a small decrease is
seen for $\cos\theta_K$ near $-1$.  The efficiency for mistagged events is maximal near $\cos\theta_K=0$,
with an increase as $\cos\theta_K$ approaches $+1$ that becomes more pronounced as $q^2$ increases.

The fit is performed in two steps.  The initial fit uses the data from the sidebands of the $\PBz$
mass to obtain the $B^{\theta_\PK}(\theta_\PK)$ and $B^{\theta_l}(\theta_l)$ distributions (the signal
component is absent from this fit).  The sideband regions are $3\sigma_m < \abs{m-m_{\PBz}} <
5.5\sigma_m$, where $\sigma_m$ is the average mass resolution ($\approx$45\MeV), obtained from
fitting the MC simulation signal to a sum of two Gaussians with a common mean.  The distributions obtained in
this step are then fixed for the second step, which is a fit to the data over the full mass range.
The free parameters in this fit are $A_\mathrm{FB}$, $F_\mathrm{L}$, $F_\mathrm{S}$,
$A_\mathrm{S}$, the parameters in $B^m(m)$, the mass parameter in $S^{C}(m)$ and $S^{M}(m)$, and
the yields $Y^{C}_{S}$ and $Y_{B}$.  In addition, the remaining parameters in $S^{C}(m)$ and
$S^{M}(m)$ are free parameters with Gaussian constraints from previous fits to simulated
signal events.

The PDF in Eq.~(\ref{eq:PDF}) is only guaranteed to be nonnegative for particular ranges of
$A_\mathrm{FB}$, $F_\mathrm{L}$, $A_\mathrm{S}$, and $F_\mathrm{S}$.  While the definition of the
precise physical region is a more complicated expression, the approximate ranges of validity are: $0<F_\mathrm{L}<1$,
$\abs{A_\mathrm{FB}}<\frac{3}{4}\left(1-F_\mathrm{L}\right)$, $0 < F_\mathrm{S} < \min{\left[
    \frac{3(1-F_\mathrm{L})}{1+3F_\mathrm{L}},1 \right]}$, and $\abs{A_\mathrm{S}}<F_\mathrm{S} +
3F_\mathrm{L}\left(1-F_\mathrm{S}\right)$.  In addition, the interference term $A_\mathrm{S}$ must vanish if
either of the two interfering components vanish.  From Ref.~\cite{Descotes-Genon:2013vna}, this constraint
is implemented as $\abs{A_\mathrm{S}} < \sqrt{12 F_\mathrm{S}(1-F_\mathrm{S})F_\mathrm{L}}R$, where $R$ is a ratio related to
the S-wave and P-wave line shapes, estimated to be 0.89 near the $\cPKstz$ mass.
During the \textsc{minuit}~\cite{Minuit} minimization,
penalty terms are introduced to ensure that parameters remain in the physical region.
When assessing the statistical uncertainties with \textsc{Minos}~\cite{Minuit}, the penalty terms
are removed.  However, a negative value for Eq.~(\ref{eq:PDF}) results in the minimizing algorithm generating
a large positive jump in the negative log-likelihood, tending to remove the unphysical region.
The results of the fit in each signal $q^2$ bin are $A_\mathrm{FB}$, $F_\mathrm{L}$, $A_\mathrm{S}$, $F_\mathrm{S}$, and the
correctly tagged signal yield $Y^{C}_S$.

The differential branching fraction, $\rd{}\mathcal{B}/\rd{}q^2$, is measured relative to the
normalization channel \BtoKstJpsi using:
\ifthenelse{\boolean{cms@external}}{
\begin{multline} \label{eq:BF}
\frac{\rd{}\mathcal{B}\left(\BtoKstmumu\right)}{\rd{}q^2} = \left(\frac{Y^{C}_{S}}{\epsilon^{C}}+\frac{Y^{C}_{S}f^{M}}{(1-f^{M})\epsilon^{M}}\right)\\
\times
\left(\frac{Y^{C}_{N}}{\epsilon^{C}_{N}}+\frac{Y^{C}_{N}f^{M}_{N}}{(1-f^{M}_{N})\epsilon^{M}_{N}}\right)^{-1}
\frac{\mathcal{B}\left(\BtoKstJpsi\right)}{\Delta q^2},
\end{multline}
}{
\begin{equation} \label{eq:BF}
\frac{\rd{}\mathcal{B}\left(\BtoKstmumu\right)}{\rd{}q^2} = \left(\frac{Y^{C}_{S}}{\epsilon^{C}}+\frac{Y^{C}_{S}f^{M}}{(1-f^{M})\epsilon^{M}}\right)
\left(\frac{Y^{C}_{N}}{\epsilon^{C}_{N}}+\frac{Y^{C}_{N}f^{M}_{N}}{(1-f^{M}_{N})\epsilon^{M}_{N}}\right)^{-1}
\frac{\mathcal{B}\left(\BtoKstJpsi\right)}{\Delta q^2},
\end{equation}
}
where $Y^{C}_{S}$ and $Y^{C}_N$ are the yields of the correctly tagged signal and normalization
channels, respectively; $\epsilon^{C}_{S}$ and $\epsilon^{C}_N$ are the efficiencies for the correctly
tagged signal and normalization channels, respectively; $f^{M}$ and $f^{M}_N$ are the mistag rates for
the signal and normalization channels, respectively; $\epsilon^{M}_{S}$ and $\epsilon^{M}_N$ are the
efficiencies for the mistagged signal and normalization channels, respectively; and
$\mathcal{B}\left(\PBz\to\cPJgy(\Pgmp \Pgmm) \cPKstz \right) = 0.132\% \times 5.96\%$ is the accepted branching 
fraction for the normalization channel~\cite{PDG}, corresponding to the $q^2$ bin
$\Delta q^2 = 8.68 - 10.09\GeV^2$.  The efficiencies are obtained by integrating the
efficiency functions over the angular variables, weighted by the decay rate in Eq.~(\ref{eq:angALL}), using
the values obtained from the fit of Eq.~(\ref{eq:PDF}) to the data.

The fit formalism and results are validated through fits to pseudo-experimental samples,
MC simulation samples, and control channels.  Additional details, including the
sizes of the systematic uncertainties assigned from these fits, are described in Section~\ref{sec:Systematics}.

\section{Systematic uncertainties}
\label{sec:Systematics}

Since the efficiency is computed with simulated events, it is essential that the MC simulation program
correctly reproduces the data, and extensive checks have been performed to verify the accuracy of the
simulation.  The systematic uncertainties associated with the efficiencies, and other
sources of systematic uncertainty are described below and summarized in Table~\ref{tab:systematics}.

The correctness of the fit function and the procedure for measuring the variables of interest are
verified in three ways.  First, a high-statistics MC sample (approximately 400 times that of the
data) is used to verify that the fitting procedure produces results consistent with the input values
to the simulation.  This MC sample includes the full simulation of signal and control channel events
plus background events obtained from the PDF in Eq.~(\ref{eq:PDF}).  The discrepancy between the
input and output values in this check is assigned as a simulation mismodeling systematic
uncertainty.
It was also verified that fitting a sample with only mistagged events gives the correct results.
Second, 1000 pseudo-experiments, each with the same number of events as the data
sample, are generated in each $q^2$ bin using the PDF in Eq.~(\ref{eq:PDF}),
with parameters obtained from the fit to the data.
These are used to estimate the fit bias.
Much of the observed bias is a consequence of the fitted parameters lying close to the boundaries of the physical region.
In addition, the distributions of results are used to check the returned
statistical uncertainty from the fit and are found to be consistent.  Third, the high-statistics MC
signal sample is divided into 400 subsamples and combined with background events to mimic 400
independent data sets of similar size to the data.  Fits to these 400 samples do not reveal any
additional systematic uncertainty.

Because the efficiency functions are estimated from a finite number of simulated events, there is a
corresponding statistical uncertainty in the efficiency.  The efficiency functions are obtained from
fits to simulated data.  Alternatives to the default efficiency function are generated by randomly
varying the fitted parameters within their uncertainties (including all correlations).  The
effect of these different efficiency functions on the final result is used to estimate
the systematic uncertainty.

The main check of the correctness of the efficiency is obtained by comparing the
efficiency-corrected results for the control channels with the corresponding world-average values.
The efficiency as a function of the angular variables is checked by comparing the $F_\mathrm{L}$ and
$A_\mathrm{FB}$ measurements from the \BtoKstJpsi sample, composed of 165\,000 signal events.
The value of $F_\mathrm{L}$ obtained in this analysis is $0.537 \pm 0.002\stat$, compared with the
world-average value of $0.571 \pm 0.007\,\text{(stat+syst)}$~\cite{PDG}, indicating a discrepancy
of $0.034$, which is taken as the systematic uncertainty for the signal measurements of
$F_\mathrm{L}$.  For $A_\mathrm{FB}$, the measured value is $0.008 \pm 0.003\stat$, compared to a
SM expectation of $\approx$0.  Adding an S-wave contribution in the
fit changes the measured value of $A_\mathrm{FB}$ by less than $0.001$.  From this, we conclude that
the S-wave effects are minimal, and assign a systematic uncertainty of $0.005$ for $A_\mathrm{FB}$.
To validate that the simulation accurately reproduces the efficiency as a function of $q^2$, we
measure the branching ratio between two different $q^2$ bins, namely the two control channels. The
branching ratio result, $\mathcal{B}\left(\BtoKstpsip\right)/\mathcal{B}\left(\BtoKstJpsi\right) = 0.479 \pm 0.005$,
is in excellent agreement with the most precise reported measurement: $0.476 \pm 0.014\stat \pm
0.010\syst$~\cite{Aaij:2012dda}.

The PDF used in the analysis accommodates cases in which the kaon and pion charges are correctly
and incorrectly assigned.  Both of these contributions are treated as signal.  The mistag
fraction is fixed to the value obtained from MC simulation.  In the high-statistics control channel
\BtoKstJpsi, the mistag fraction is allowed to float in the fit and a value of $f^{M} = (14.5 \pm 0.5)\%$ is
found, to be compared to the simulated value of $(13.7 \pm 0.1)\%$.  The effect of this 5.8\%
difference in the mistag fraction on the measured values is taken as a systematic uncertainty.

The systematic uncertainty associated with the functions used to model the angular distribution of
the background is obtained from the sum in quadrature of two uncertainties.  The first uncertainty
is evaluated by fitting the background with polynomials of one degree greater than used in the
default analysis and taking the difference in the observables of interest between these two fits as
the systematic uncertainty.  The second uncertainty is owing to the statistical uncertainty in the
background shape, as these shapes are fixed in the final fit.  This uncertainty is obtained by taking the
difference in quadrature between the returned statistical uncertainties on the parameters of interest
when the background shapes are fixed and allowed to vary.  In $q^2$ bins where the unconstrained fit
does not converge, the associated uncertainty is obtained from extrapolation of nearby bins.

The mass distributions for the correctly tagged and mistagged events are each described by the sum of two
Gaussian functions, with a common mean for all four Gaussian functions.  The mean value is obtained
from the fit to the data, while the other parameters (four $\sigma$ and two ratios) are
obtained from fits to MC-simulated events, with the uncertainty from those fits used as Gaussian
constraints in the fits to the data.  For the high-statistics control channels, it
is possible to fit the data, while allowing some of the parameters to vary.  The maximum
changes in the measured values in the two control channel $q^2$ bins when the parameters
are varied are taken as the systematic uncertainty for all $q^2$ bins.

The $q^2$ bins just below and above the \cPJgy\ region may be contaminated with \BtoKstJpsi
feed-through events that are not removed by the selection criteria.  A special fit in these
two bins is made, in which an additional background term is added to the PDF\@.  This background
distribution is obtained from the MC simulation and the background yield is a free parameter.  The
resulting changes in the fit parameters are used as estimates of the systematic uncertainty
associated with this contribution.

The effects from angular resolution in the reconstructed values for the angular variables
$\theta_\PK$ and $\theta_l$ are estimated by performing two fits on the same MC-simulated
events.  One fit uses the true values of the angular variables and the other fit their
reconstructed values.  The difference in the fitted parameters between the two fits is
taken as an estimate of the systematic uncertainty.

The differential branching fraction has an additional systematic uncertainty of 4.6\% coming from the
uncertainty in the branching fraction of the normalization mode \BtoKstJpsi.

The systematic uncertainties are measured and applied in each $q^2$ bin, with the total systematic
uncertainty obtained by adding the individual contributions in quadrature.

\begin{table*}[htbp!]
  \centering
  \topcaption{\label{tab:systematics}Systematic uncertainty contributions for the measurements of
    $F_\mathrm{L}$, $A_\mathrm{FB}$, and the branching fraction for the decay \BtoKstmumu.
    The values for $F_\mathrm{L}$ and $A_\mathrm{FB}$ are absolute, while the values for the branching
    fraction are relative.  The total uncertainty in each $q^2$ bin is obtained by adding
    each contribution in quadrature.  For each item, the range indicates the variation of the uncertainty
    in the signal $q^2$ bins.}
\begin{tabular}{l|cccc}
Systematic uncertainty & $F_\mathrm{L} (10^{-3})$  & $A_\mathrm{FB} (10^{-3})$ & $\rd{}\mathcal{B}/\rd{}q^2$ (\%)\\[1pt]
\hline\\[-2ex]
Simulation mismodeling       & 1--17  & 0--37  & 1.0--5.5 \\[1pt]
Fit bias                     & 0--34  & 2--42  & --- \\[1pt]
MC statistical uncertainty   & 3--10  & 5--18  & 0.5--2.0 \\[1pt]
Efficiency                   & 34     & 5      & --- \\[1pt]
$\PK\pi$ mistagging          & 1--4   & 0--7   & 0.1--4.1 \\[1pt]
Background distribution      & 20--36 & 12--31 & 0.0--1.2 \\[1pt]
Mass distribution            & 3      & 1      & 3.2 \\[1pt]
Feed-through background      & 0--27  & 0--5   & 0.0--4.0 \\[1pt]
Angular resolution           & 6--24  & 0--5   & 0.2--2.1 \\[1pt]
Normalization to \BtoKstJpsi & ---    & ---    & 4.6 \\[1pt]
\hline
Total systematic uncertainty & 41--65 & 18--74 & 6.4--8.6 \\[1pt]
\end{tabular}
\end{table*}

\section{Results}
\label{sec:Results}

\begin{figure*}[htbp]
  \begin{center}
    \includegraphics[width=0.49\textwidth]{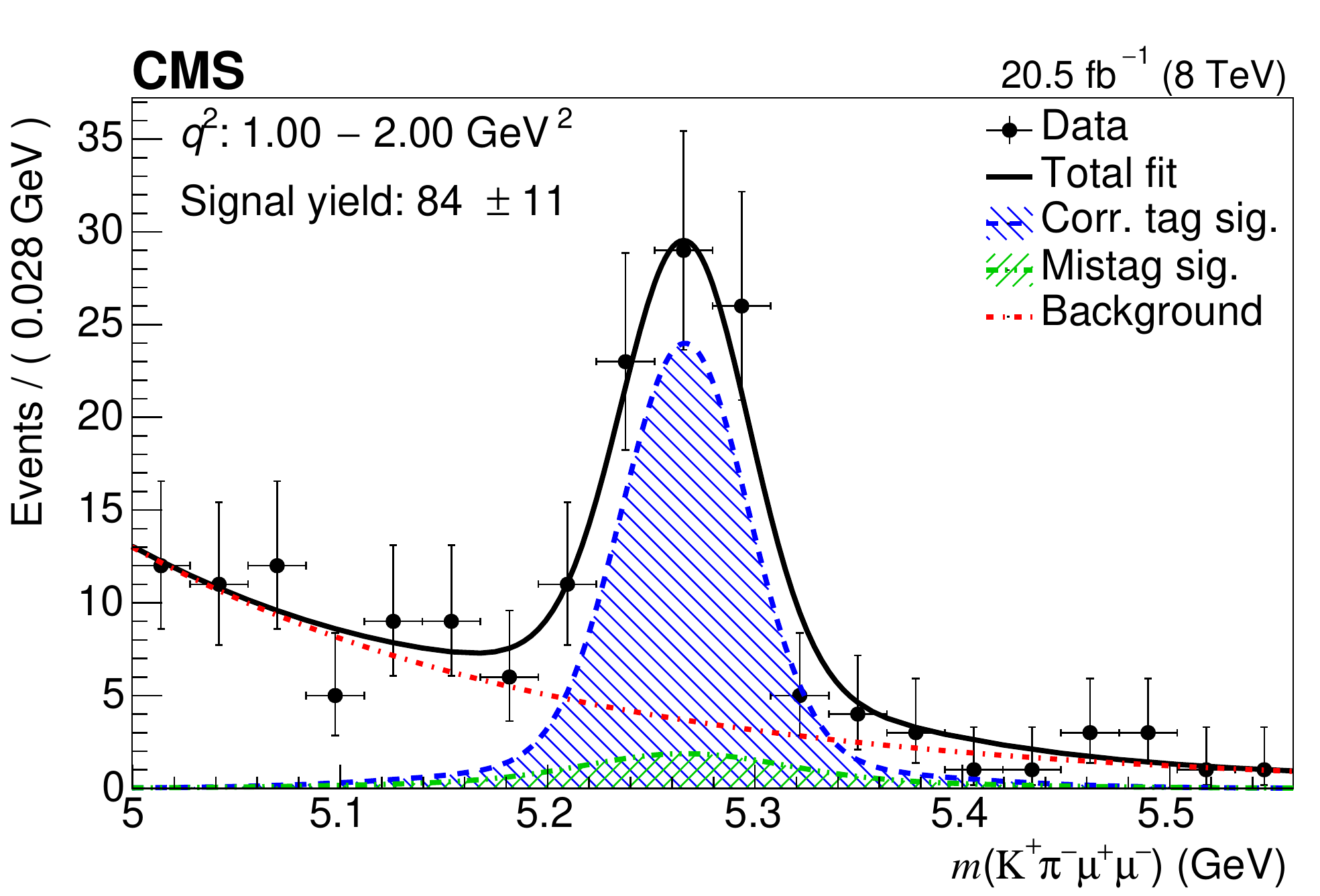}\hspace{5pt}\includegraphics[width=0.49\textwidth]{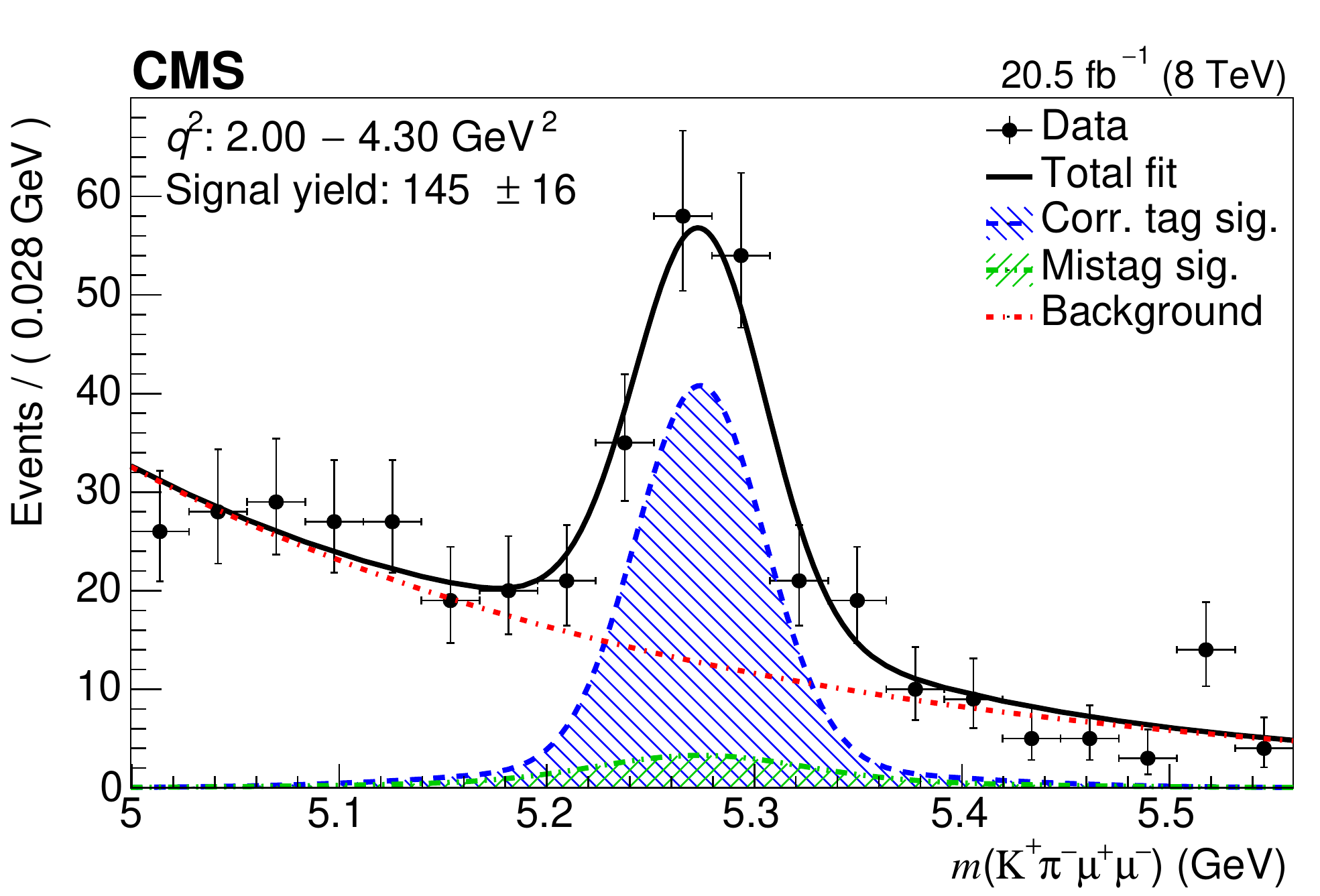}
    \includegraphics[width=0.49\textwidth]{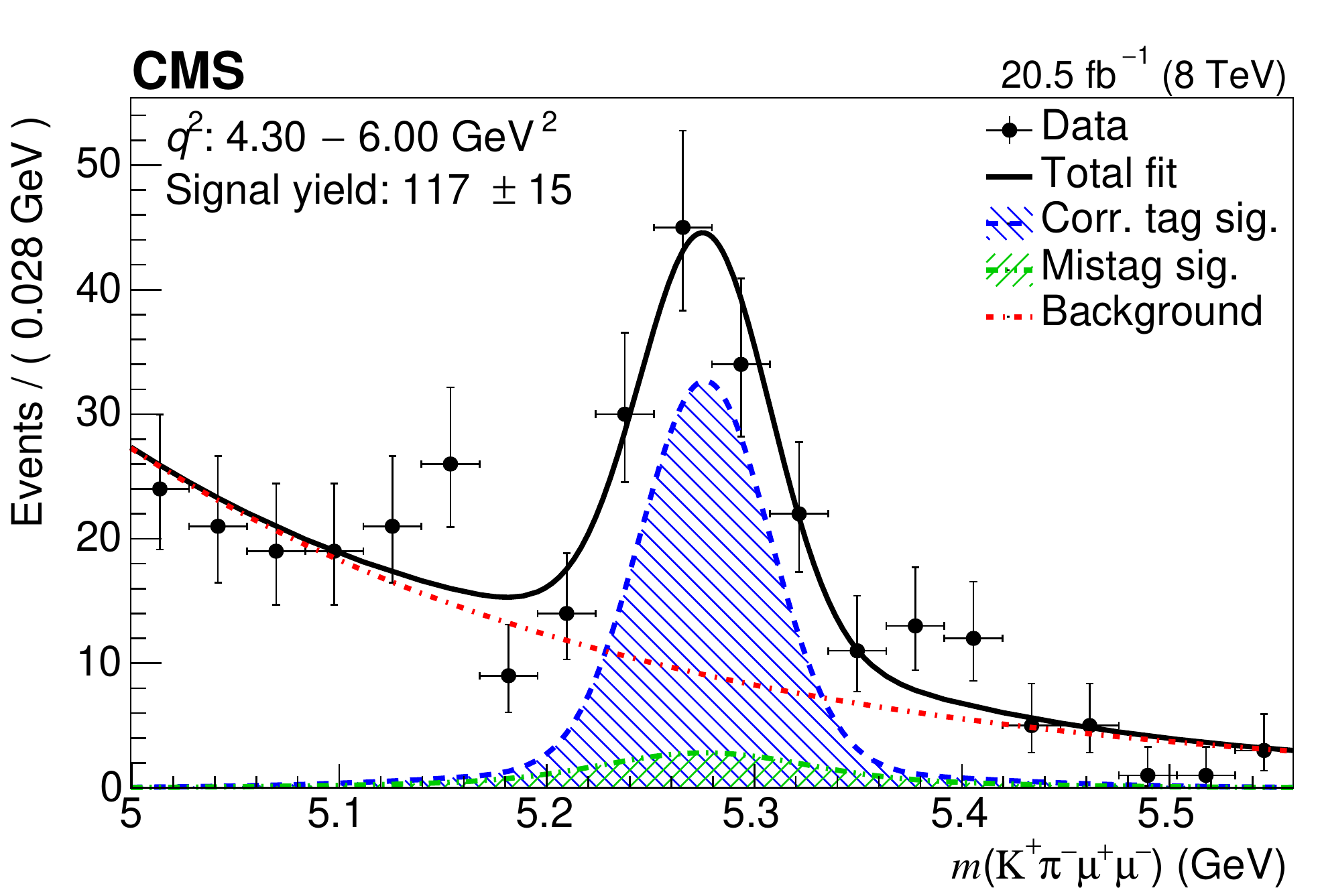}\hspace{5pt}\includegraphics[width=0.49\textwidth]{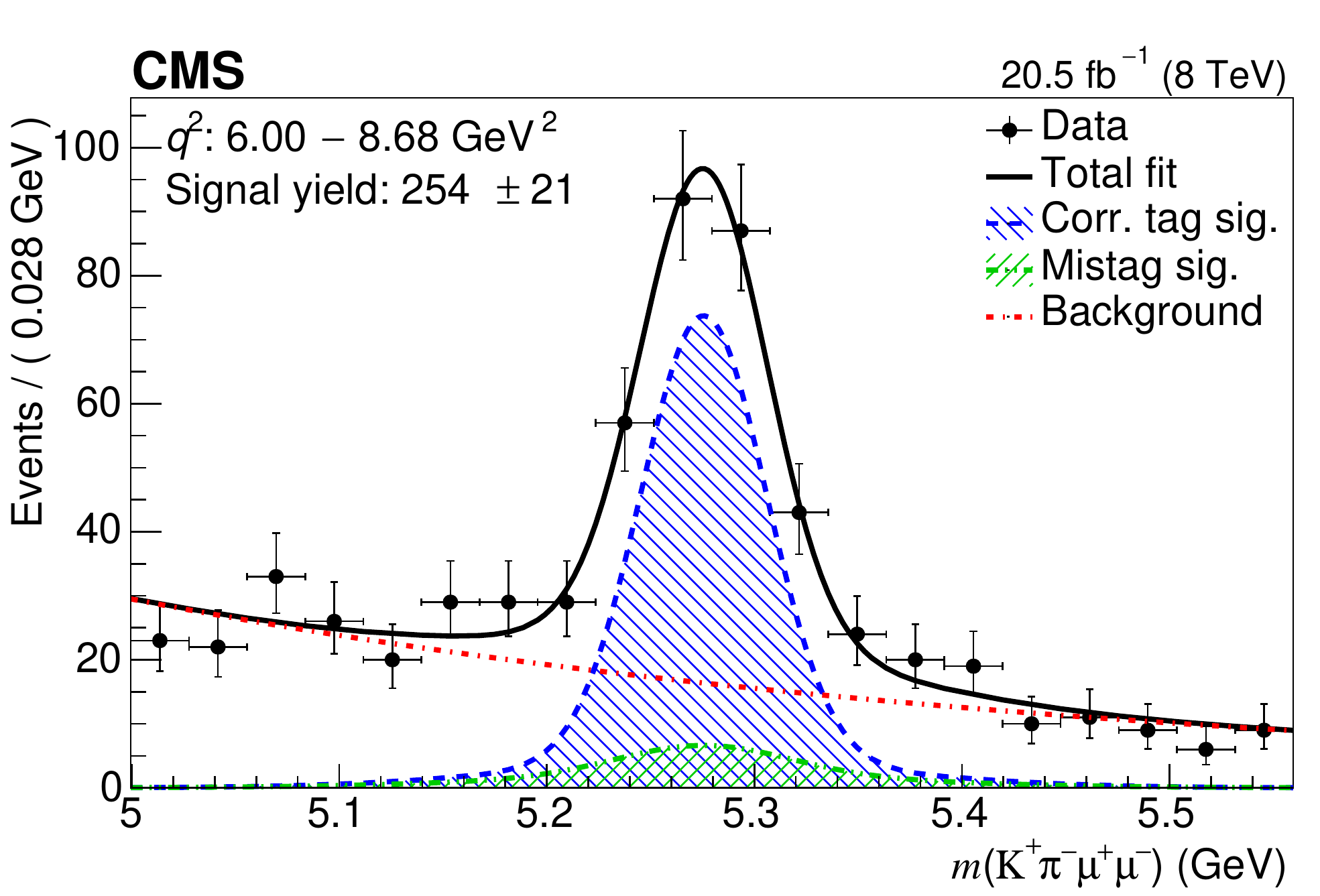}
    \includegraphics[width=0.49\textwidth]{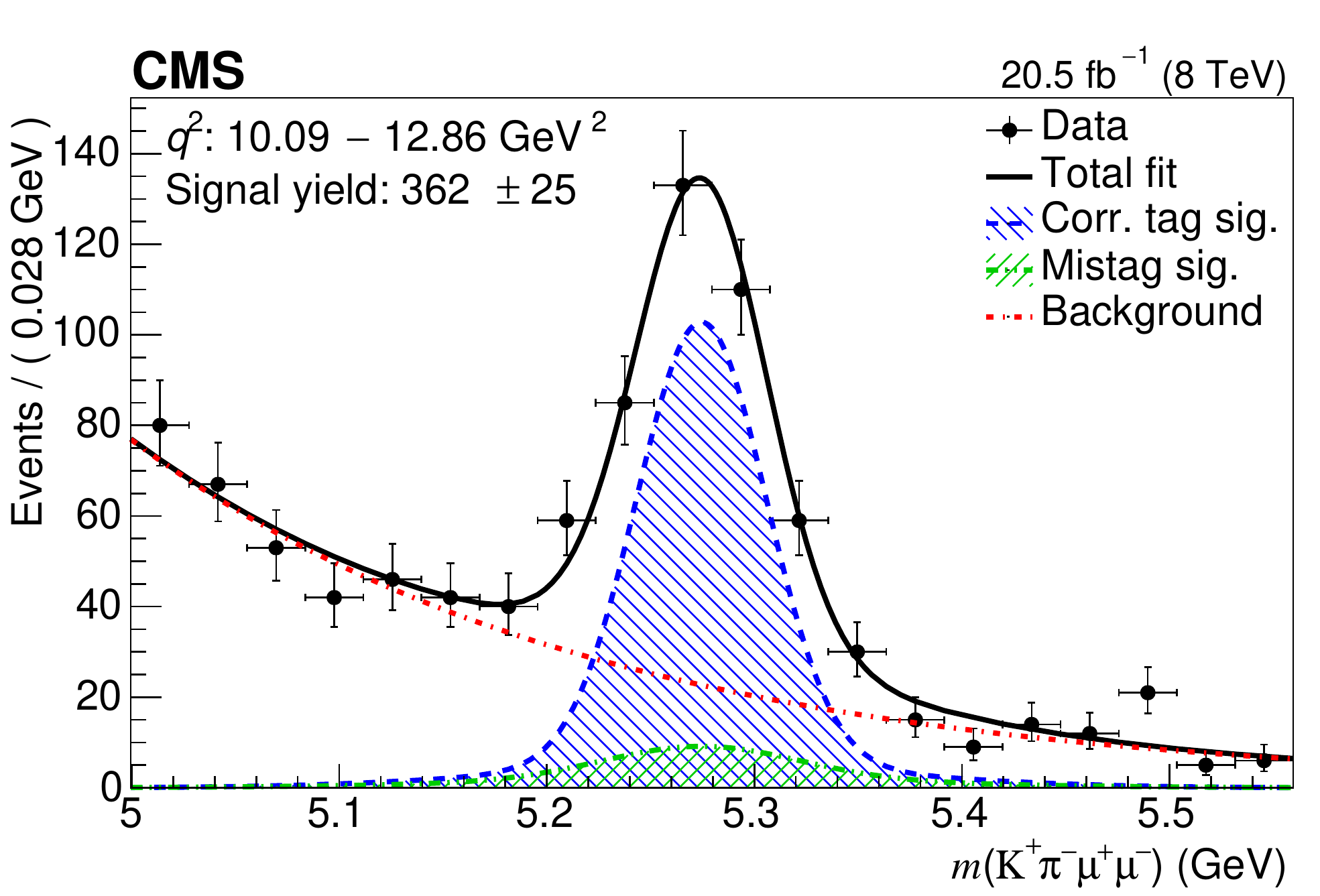}\hspace{5pt}\includegraphics[width=0.49\textwidth]{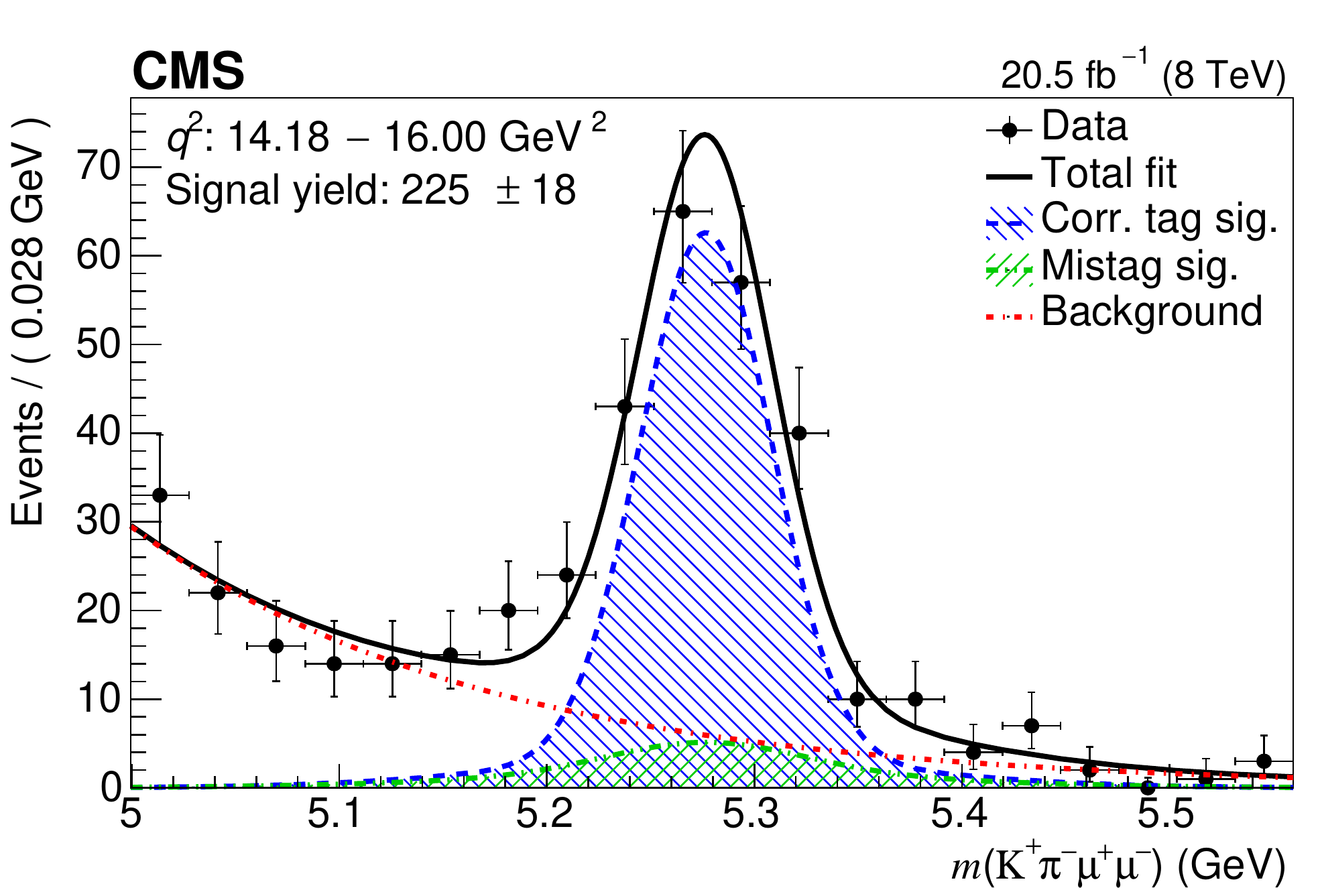}
    \includegraphics[width=0.49\textwidth]{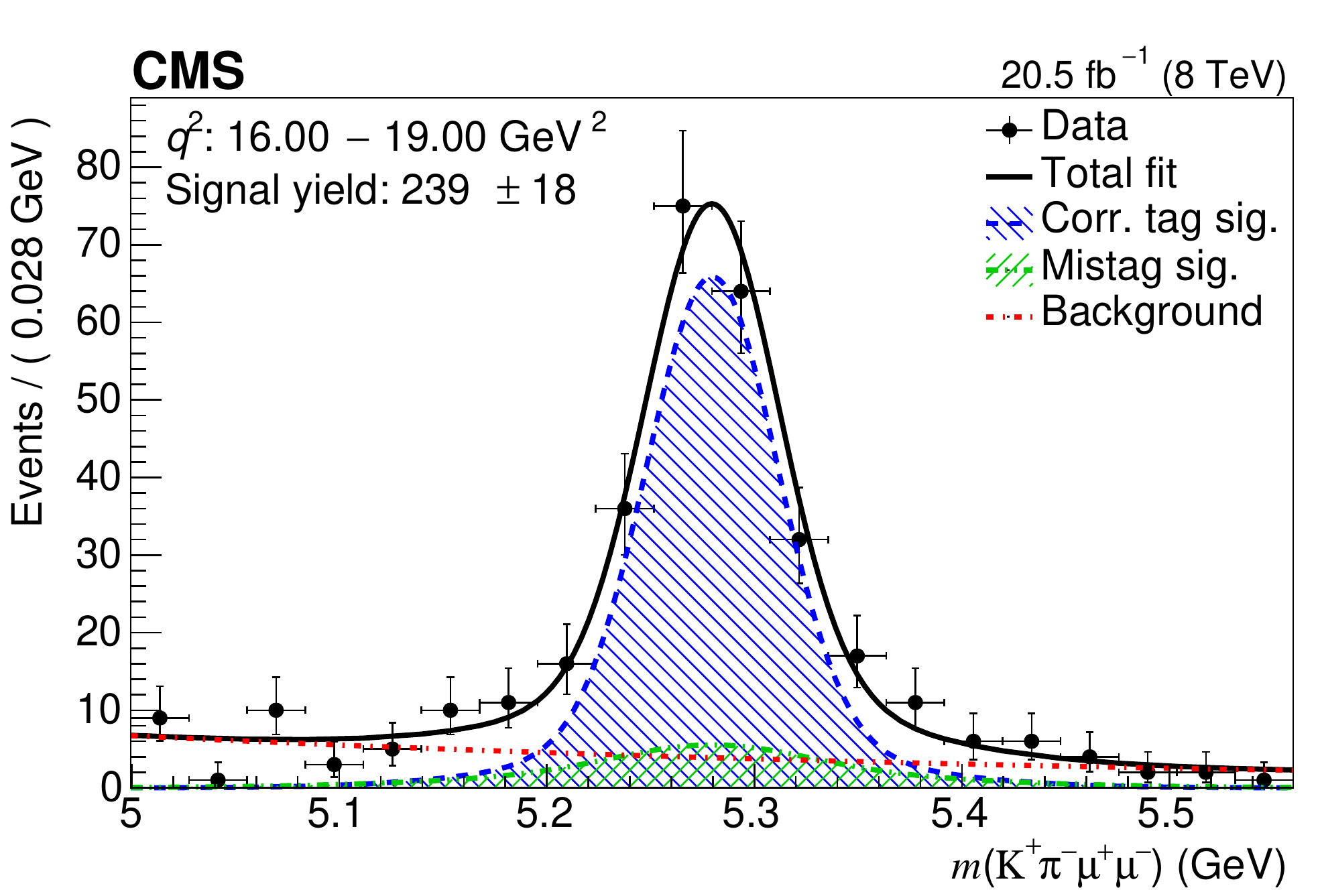}\hspace{5pt}\includegraphics[width=0.49\textwidth]{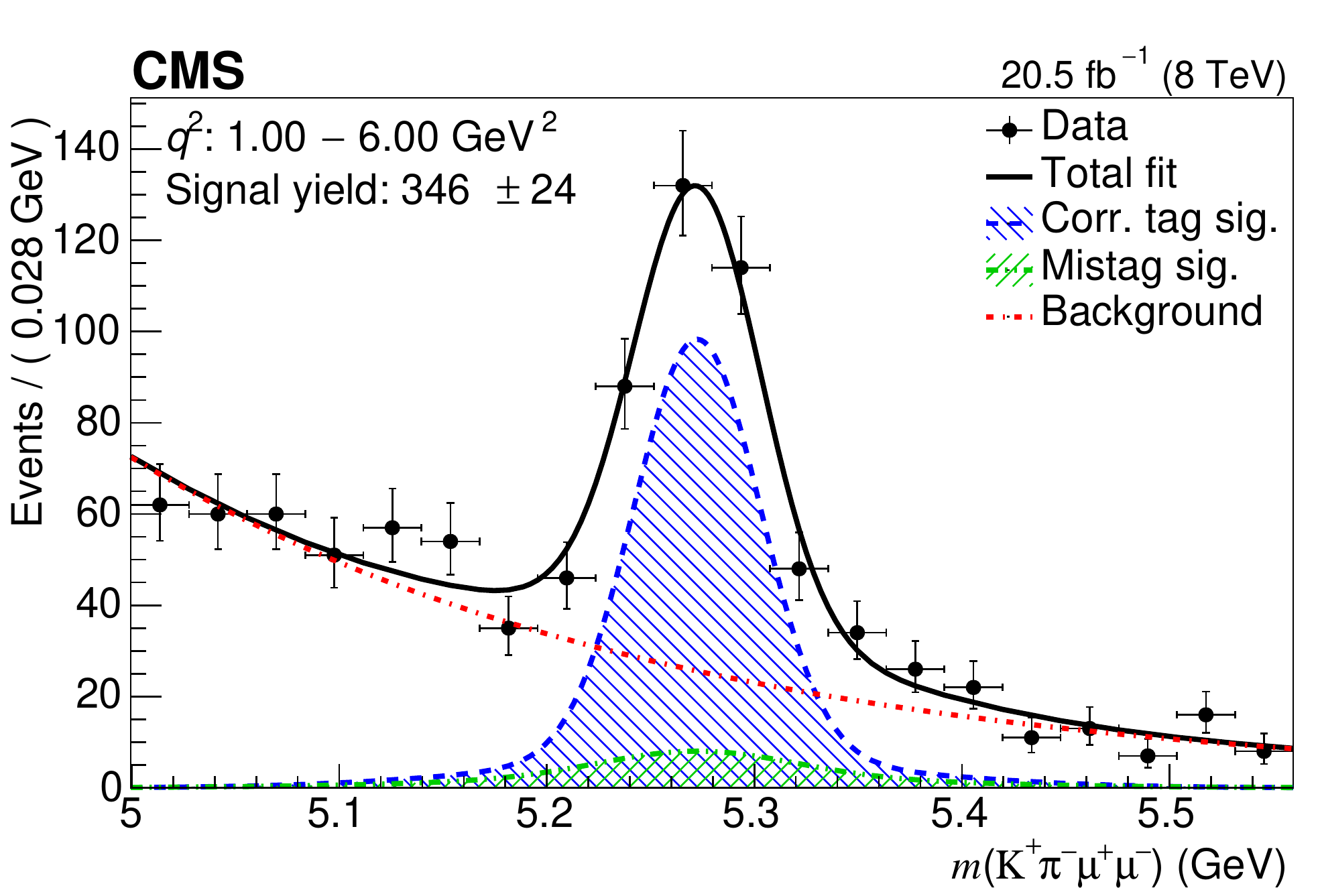}
    \caption{The $\PKp\Pgpm\Pgmp\Pgmm$ invariant mass distributions for the seven signal $q^2$ bins
      and the combined $1<q^2<6\GeV^2$ bin.  Overlaid on each is the projection of the results for
      the total fit, as well as the three components: correctly tagged signal, mistagged signal, and
      background.  The vertical bars give the statistical uncertainties, the horizontal bars the bin widths.}
    \label{fig:massplots}
  \end{center}
\end{figure*}

\begin{figure*}[htb]
  \begin{center}
    \includegraphics[width=0.49\textwidth]{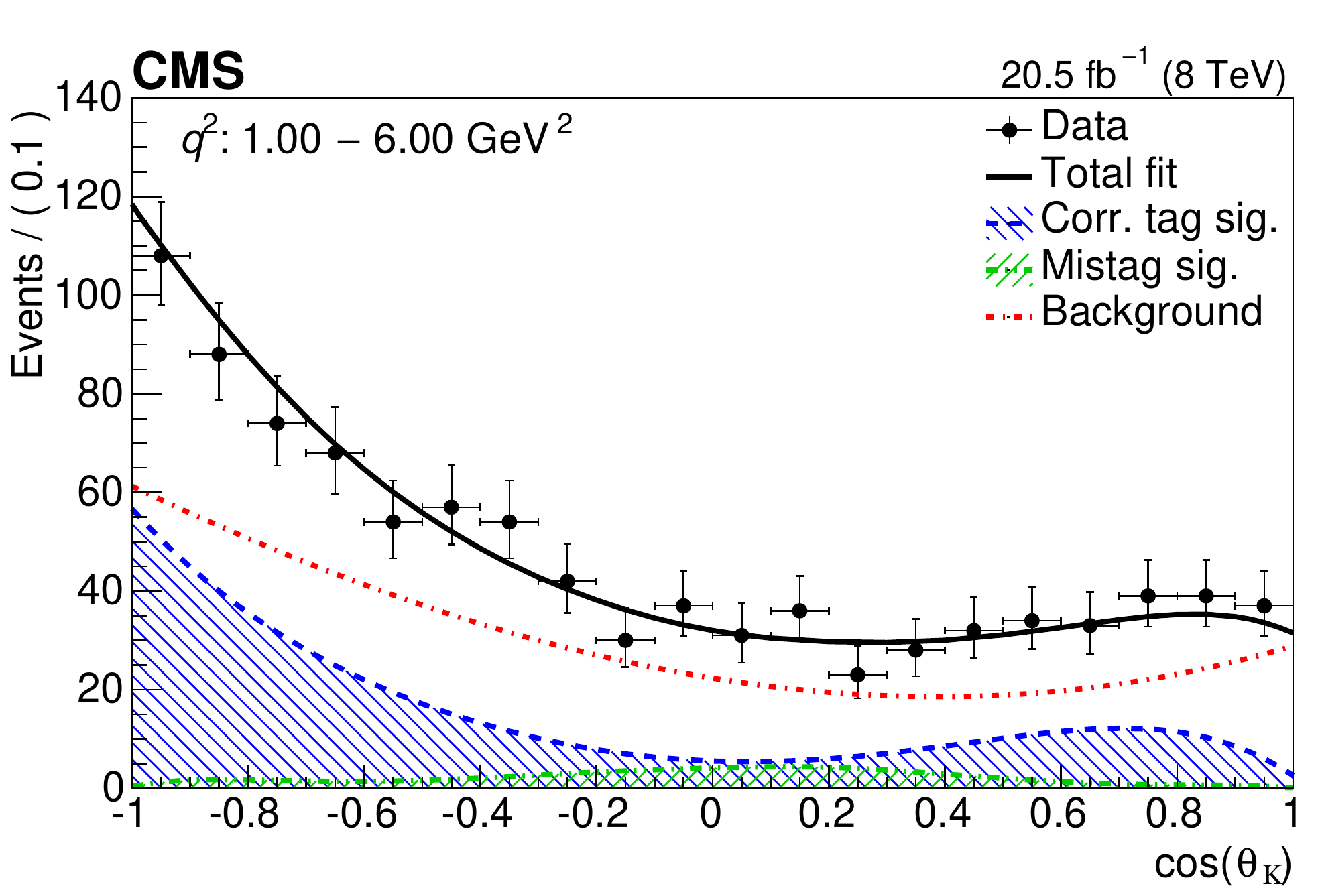}\hspace{5pt}\includegraphics[width=0.49\textwidth]{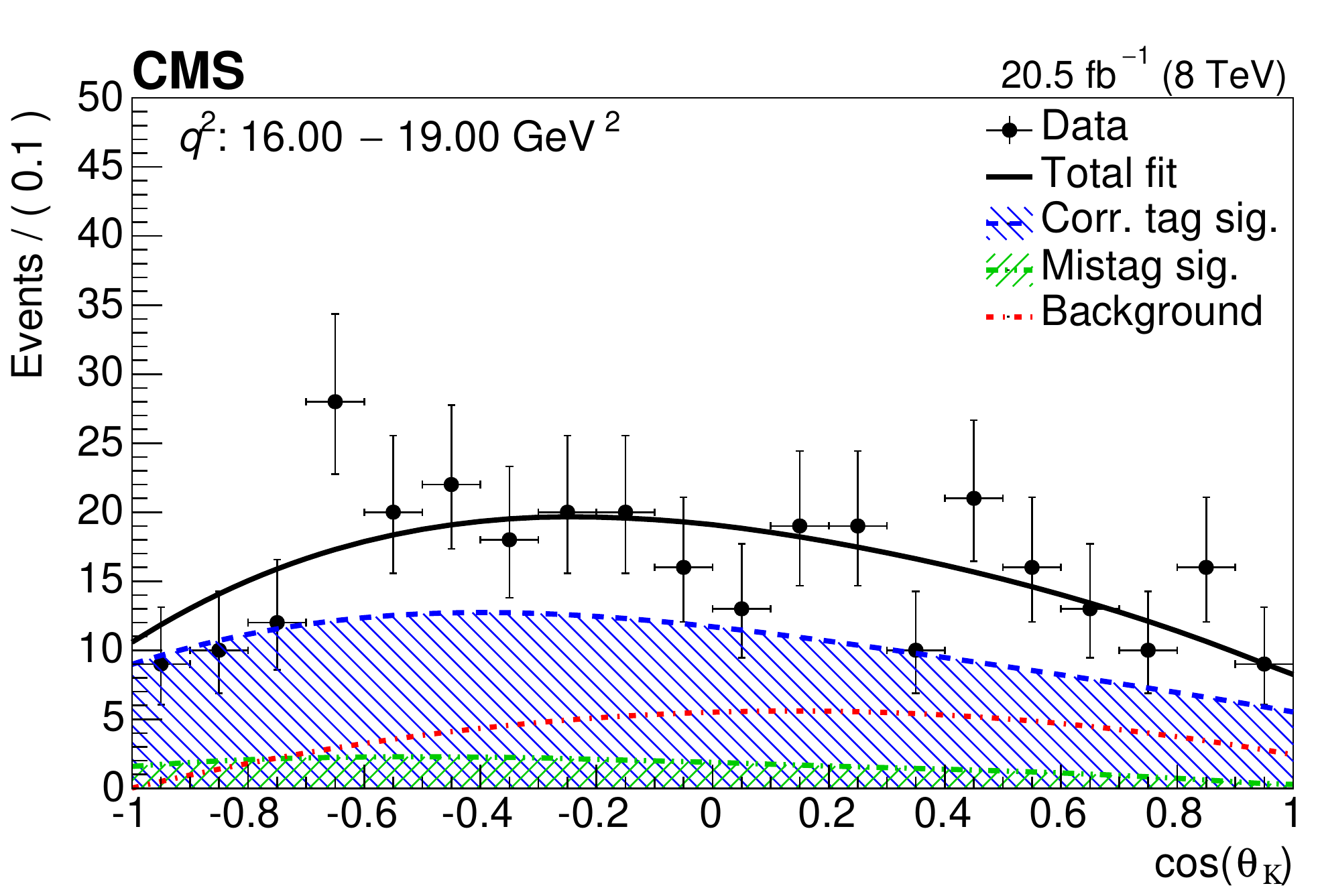}
    \includegraphics[width=0.49\textwidth]{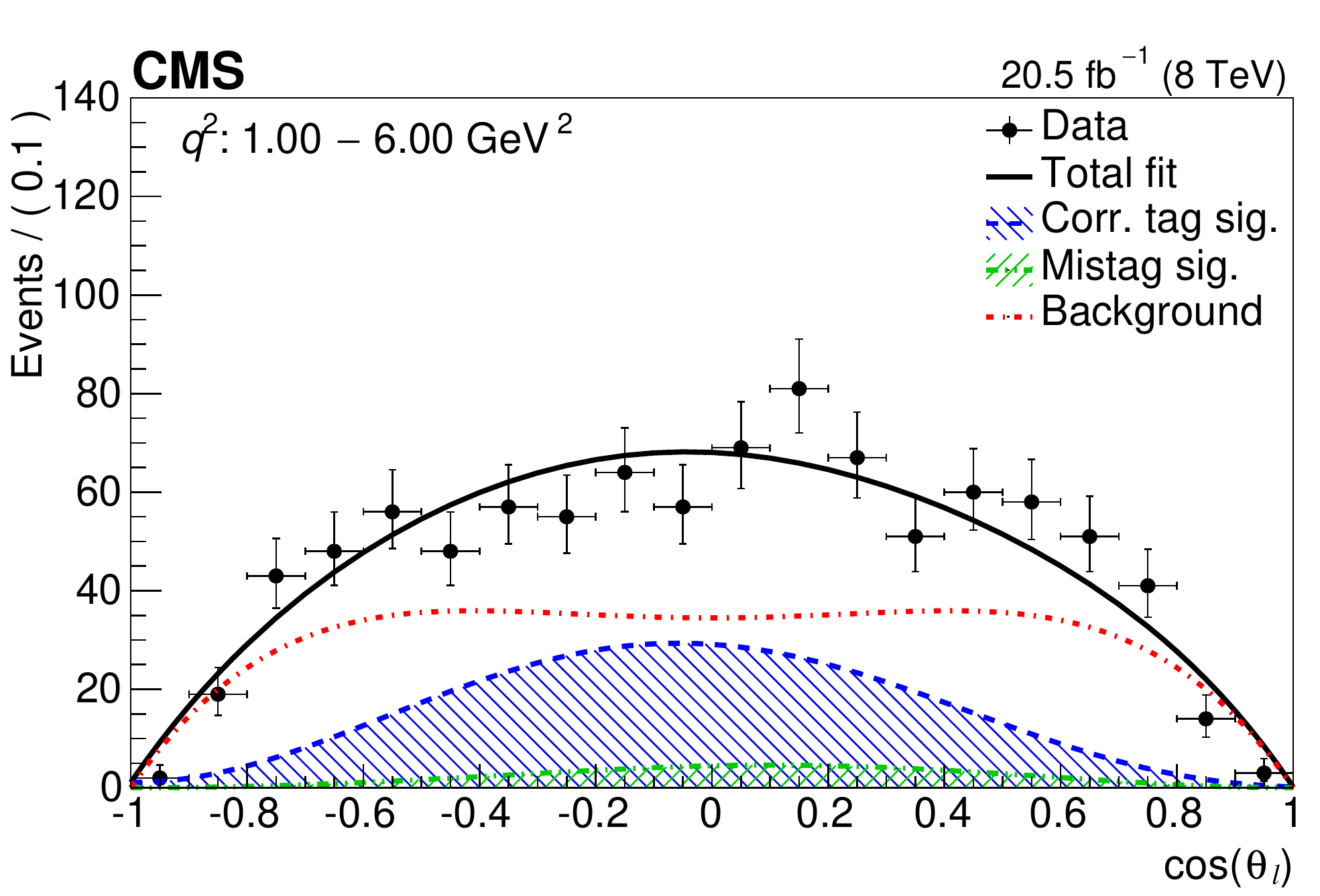}\hspace{5pt}\includegraphics[width=0.49\textwidth]{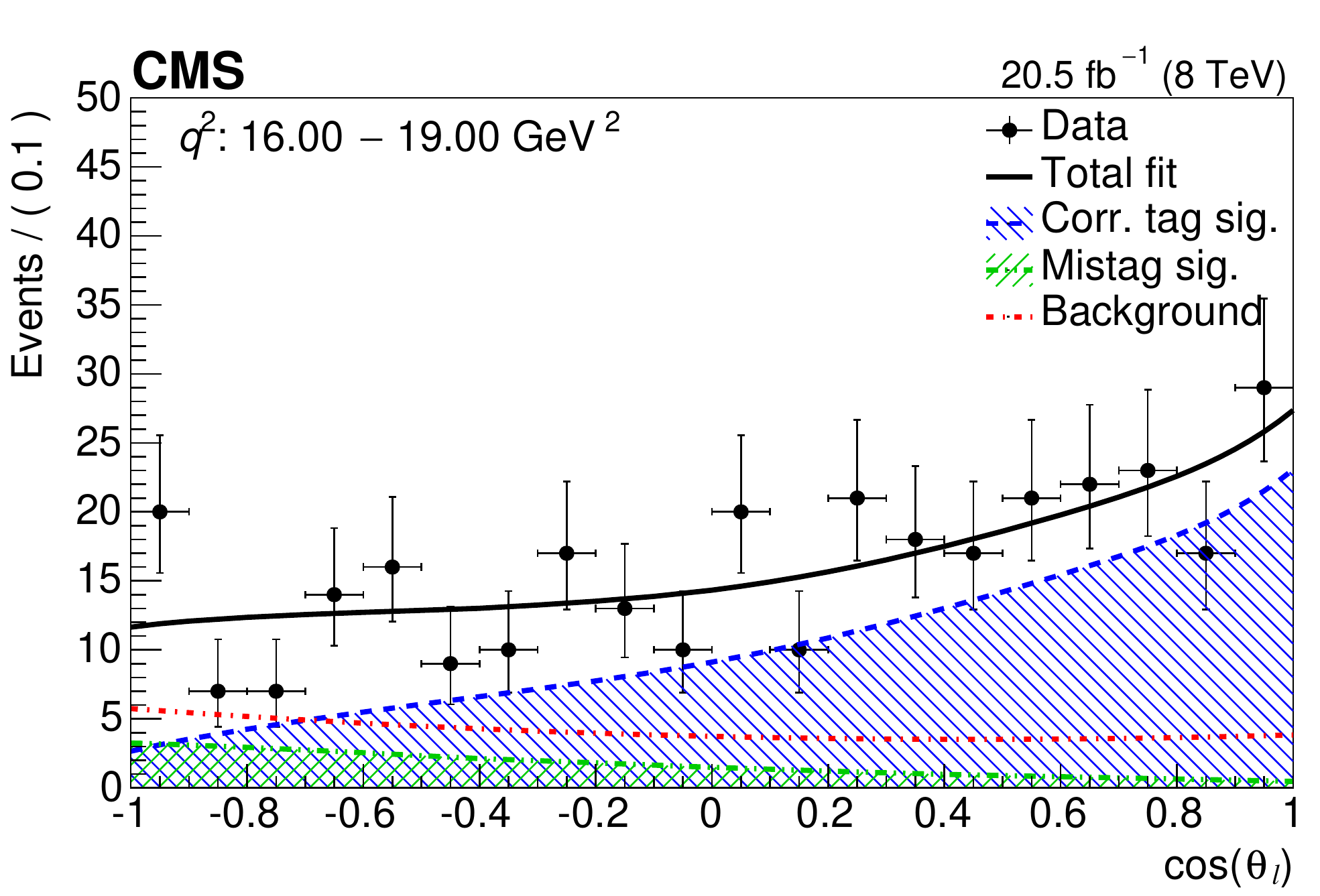}
    \caption{Data and fit results for $1<q^2<6\GeV^2$ (left) and $16<q^2<19\GeV^2$ (right),
      projected onto the $\cos\theta_\PK$ axis (top), and $\cos\theta_l$ axis (bottom).  The fit
      results show the total fit, as well as the three components: correctly tagged signal, mistagged
      signal, and background.  The vertical bars give the statistical uncertainties, the horizontal bars the bin widths.}
    \label{fig:fitprojections}
  \end{center}
\end{figure*}

\begin{figure}[htbp!]
  \begin{center}
    \includegraphics[width=0.49\textwidth]{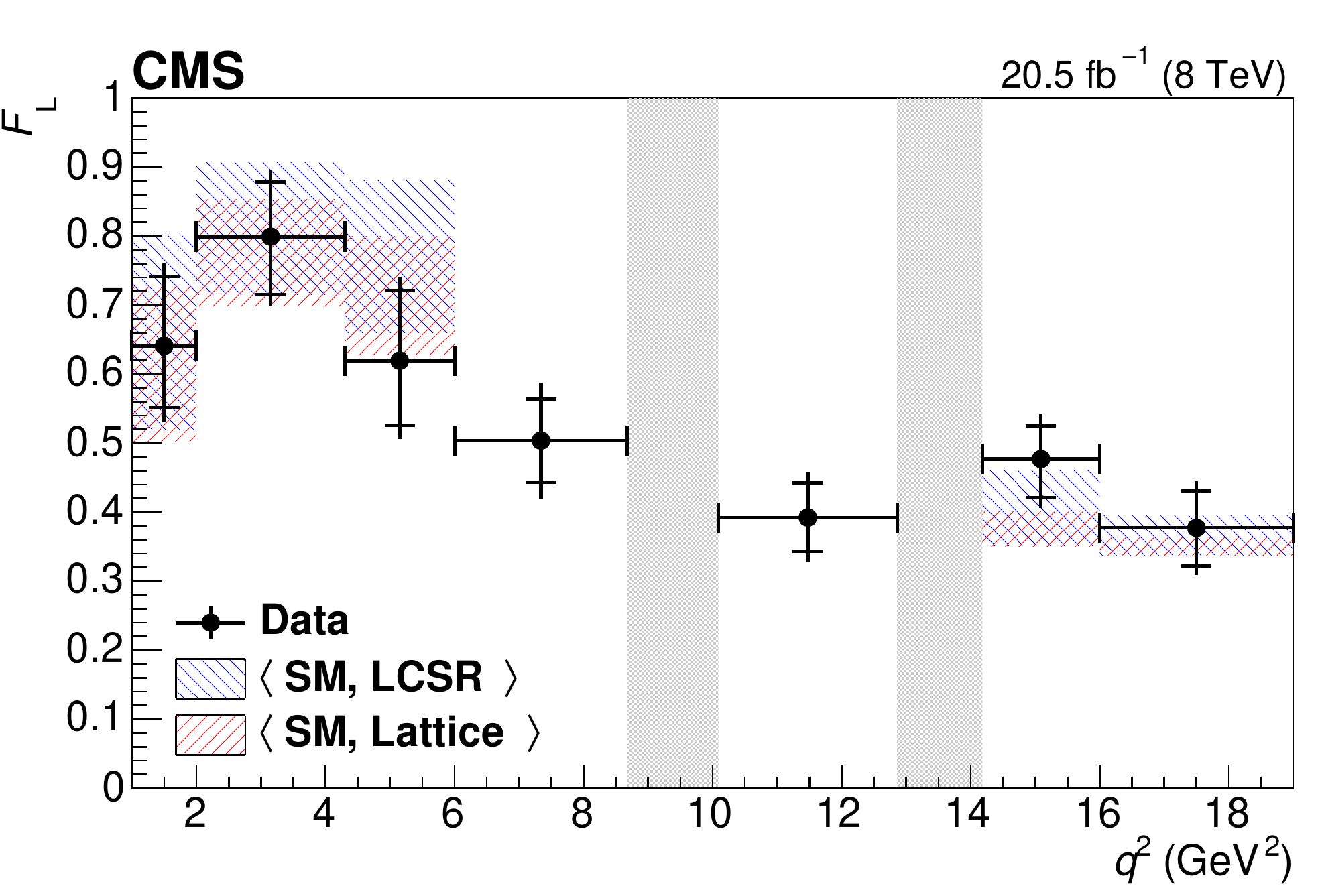}\hspace{5pt}
    \includegraphics[width=0.49\textwidth]{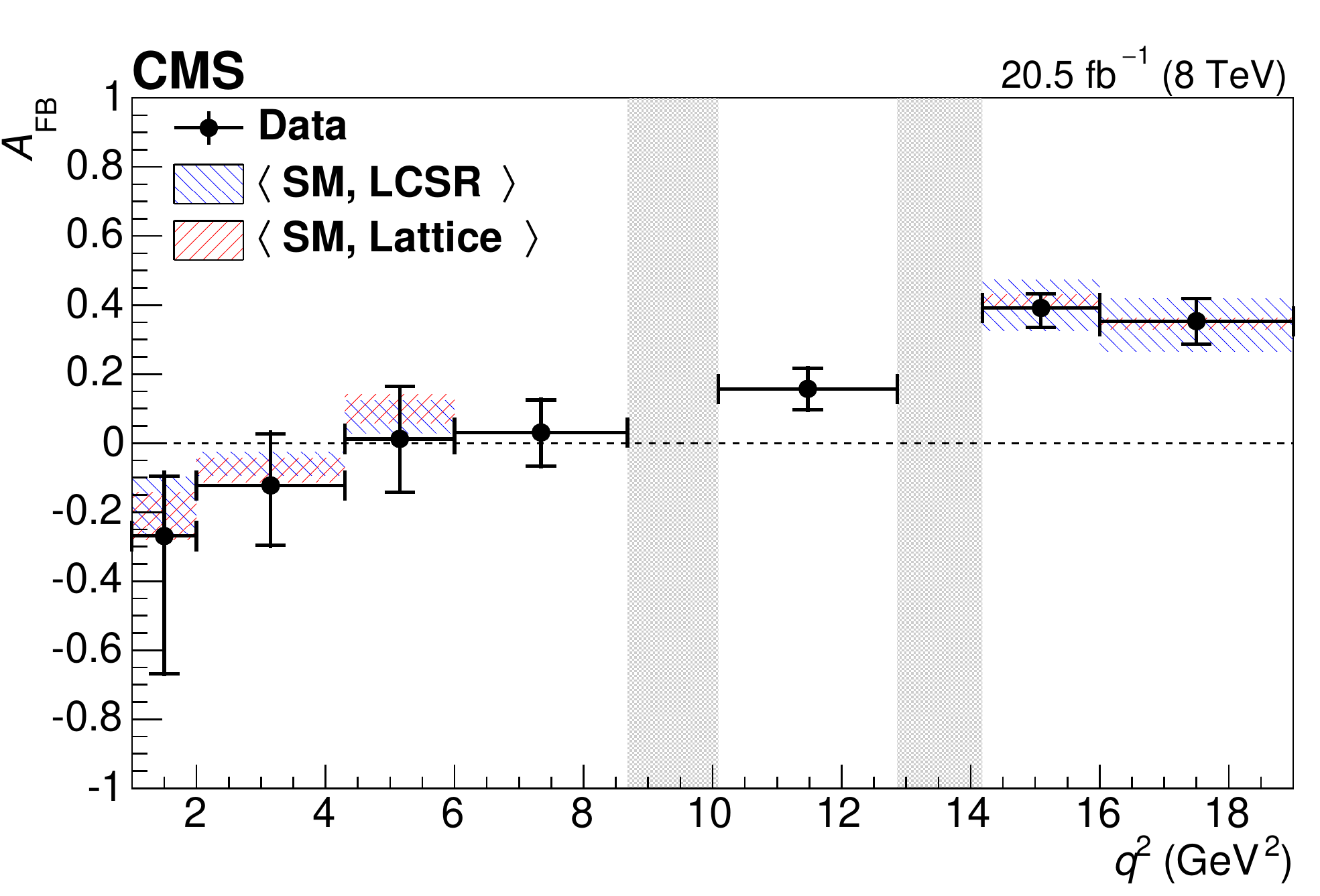}
    \includegraphics[width=0.49\textwidth]{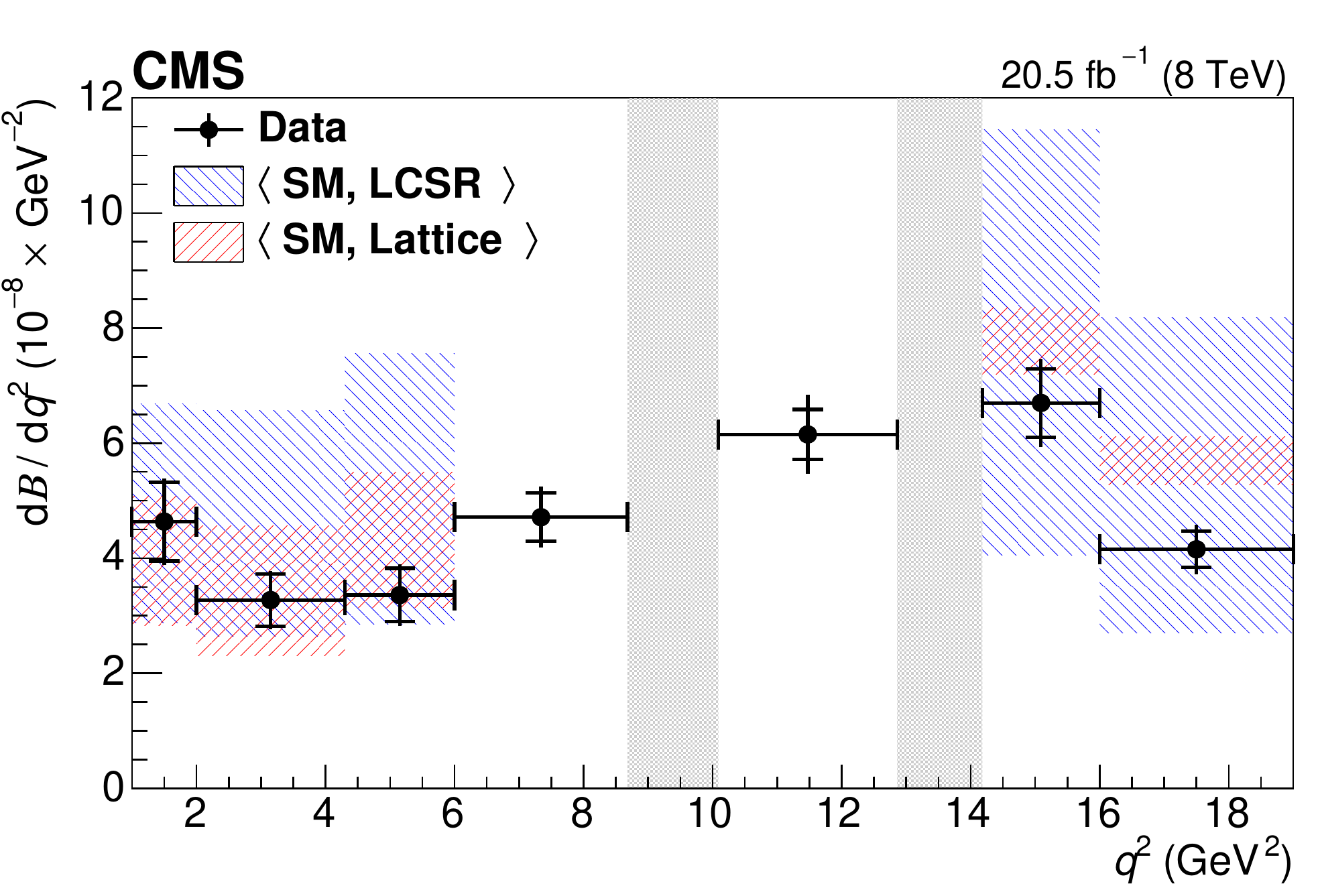}
    \caption{Measured values of $F_\mathrm{L}$, $A_\mathrm{FB}$, and $\rd{}\mathcal{B}/\rd{}q^2$ versus $q^2$
      for \BtoKstmumu.  The statistical uncertainty is shown by the inner vertical bars, while
      the outer vertical bars give the total uncertainty.  The horizontal bars show the bin widths.  The vertical shaded regions correspond to
      the $\cPJgy$ and $\psi'$ resonances.  The other shaded regions show the two SM predictions after
      rate averaging across the $q^2$ bins to provide a direct comparison to the data.  Controlled
      theoretical predictions are not available near the $\cPJgy$ and $\psi'$ resonances.}
    \label{fig:results}
  \end{center}
\end{figure}

The signal data, corresponding to 1430 signal events, are fit in seven disjoint $q^2$ bins from
1 to $19\GeV^2$.  Results are also obtained for a wide, low-$q^2$ bin $(1<q^2<6\GeV^2)$,
where the theoretical uncertainties are best understood.  The $\PKp\Pgpm\Pgmp\Pgmm$ invariant mass
distributions for all of the $q^2$ signal bins, as well as the fit projections, are shown in
Fig.~\ref{fig:massplots}.  Figure~\ref{fig:fitprojections} plots the projections of the fit and the
data on the $\cos\theta_\PK$ (top) and $\cos\theta_l$ (bottom) axes for the combined low-$q^2$ bin
(left, $1<q^2<6\GeV^2$) and the highest $q^2$ bin (right, $16<q^2<19\GeV^2$). The fitted values of
signal yield, $F_\mathrm{L}$, $A_\mathrm{FB}$, and $\rd{}\mathcal{B}/\rd{}q^2$, along with their associated
uncertainties, are given for each of the disjoint $q^2$ regions in
Table~\ref{tab:results}.  These results are also shown in Fig.~\ref{fig:results}, along with two
SM predictions.  The fitted values for $F_\mathrm{S}$ are all less than 0.03, while the values for
$A_\mathrm{S}$ vary from $-0.3$ to $+0.3$.

\begin{table*}[htb]
  \centering
  \topcaption{\label{tab:results}The measured values of signal yield (including both correctly tagged and mistagged events),
$F_\mathrm{L}$, $A_\mathrm{FB}$, and differential branching
    fraction for the decay \BtoKstmumu in bins of $q^2$.  The first uncertainty is statistical and
    the second (when present) is systematic.  The bin ranges are selected to allow comparisons to previous measurements.}
\begin{tabular}{c|ccccc}
$q^2$ & Signal & $F_\mathrm{L}$  & $A_\mathrm{FB}$ & $\rd{}\mathcal{B}/\rd{}q^2$\\
$(\GeVns^2)$ & yield &  &     & $(10^{-8}\GeV^{-2})$ \\[1pt]
\hline\\[-2ex]
1.00--2.00        & $ 84 \pm 11$ & $0.64 ^{\:+\:0.10}_{\:-\:0.09} \pm 0.07$ & $-0.27^{\:+\:0.17}_{\:-\:0.40} \pm 0.07$ & $4.6 \pm 0.7 \pm 0.3$ \\[1pt]
2.00--4.30        & $145 \pm 16$ & $0.80 \pm 0.08 \pm 0.06$              & $-0.12^{\:+\:0.15}_{\:-\:0.17} \pm 0.05$ & $3.3 \pm 0.5 \pm 0.2$ \\[1pt]
4.30--6.00        & $117 \pm 15$ & $0.62 ^{\:+\:0.10}_{\:-\:0.09} \pm 0.07$ & $ 0.01 \pm 0.15 \pm 0.03$             & $3.4 \pm 0.5 \pm 0.3$ \\[1pt]
6.00--8.68        & $254 \pm 21$ & $0.50 \pm 0.06 \pm 0.06$              & $ 0.03 \pm 0.10 \pm 0.02$             & $4.7 \pm 0.4 \pm 0.3$ \\[1pt]
10.09--12.86      & $362 \pm 25$ & $0.39 \pm 0.05 \pm 0.04$              & $ 0.16 \pm 0.06 \pm 0.01$             & $6.2 \pm 0.4 \pm 0.5$ \\[1pt]
14.18--16.00      & $225 \pm 18$ & $0.48 ^{\:+\:0.05}_{\:-\:0.06} \pm 0.04$ & $ 0.39^{\:+\:0.04}_{\:-\:0.06} \pm 0.01$ & $6.7 \pm 0.6 \pm 0.5$ \\[1pt]
16.00--19.00      & $239 \pm 18$ & $0.38 ^{\:+\:0.05}_{\:-\:0.06} \pm 0.04$ & $ 0.35 \pm 0.07 \pm 0.01$             & $4.2 \pm 0.3 \pm 0.3$ \\[1pt]
\hline
\end{tabular}
\end{table*}

The SM predictions, derived from Refs.~\cite{Bobeth:2010wg,Bobeth:2012vn}, combine two calculational
techniques. In the low-$q^2$ region, a quantum chromodynamic factorization approach~\cite{Beneke:2001at} is used,
which is applicable for $q^2<4m_c^2$, where $m_c$ is the charm quark mass.
In the high-$q^2$ region, an operator product expansion in the inverse \cPqb\ quark mass and
$1/\!\sqrt{q^2}$~\cite{Grinstein:2004vb,Beylich:2011aq} is combined with heavy-quark form-factor
relations~\cite{Grinstein:2002cz}.  This is valid above the open-charm threshold $(q^2 \gtrsim 13.9\GeV^2)$.  The two SM
predictions shown in Fig.~\ref{fig:results} differ in the
calculation of the form factors.  The light-cone sum rules (LCSR) calculation is made at
low $q^2$~\cite{Khodjamirian:2010} and is extrapolated to high $q^2$~\cite{Khodjamirian:2012rm}.
The lattice gauge (Lattice) calculation of the form factors is from Ref.~\cite{Horgan:2013hoa}.
Controlled theoretical predictions are not available near the $\cPJgy$ and $\psi'$ resonances.
The SM predictions are in good agreement with the CMS experimental results, indicating no strong
contribution from physics beyond the standard model.

\begin{figure}[!htbp]
  \begin{center}
    \includegraphics[width=0.49\textwidth]{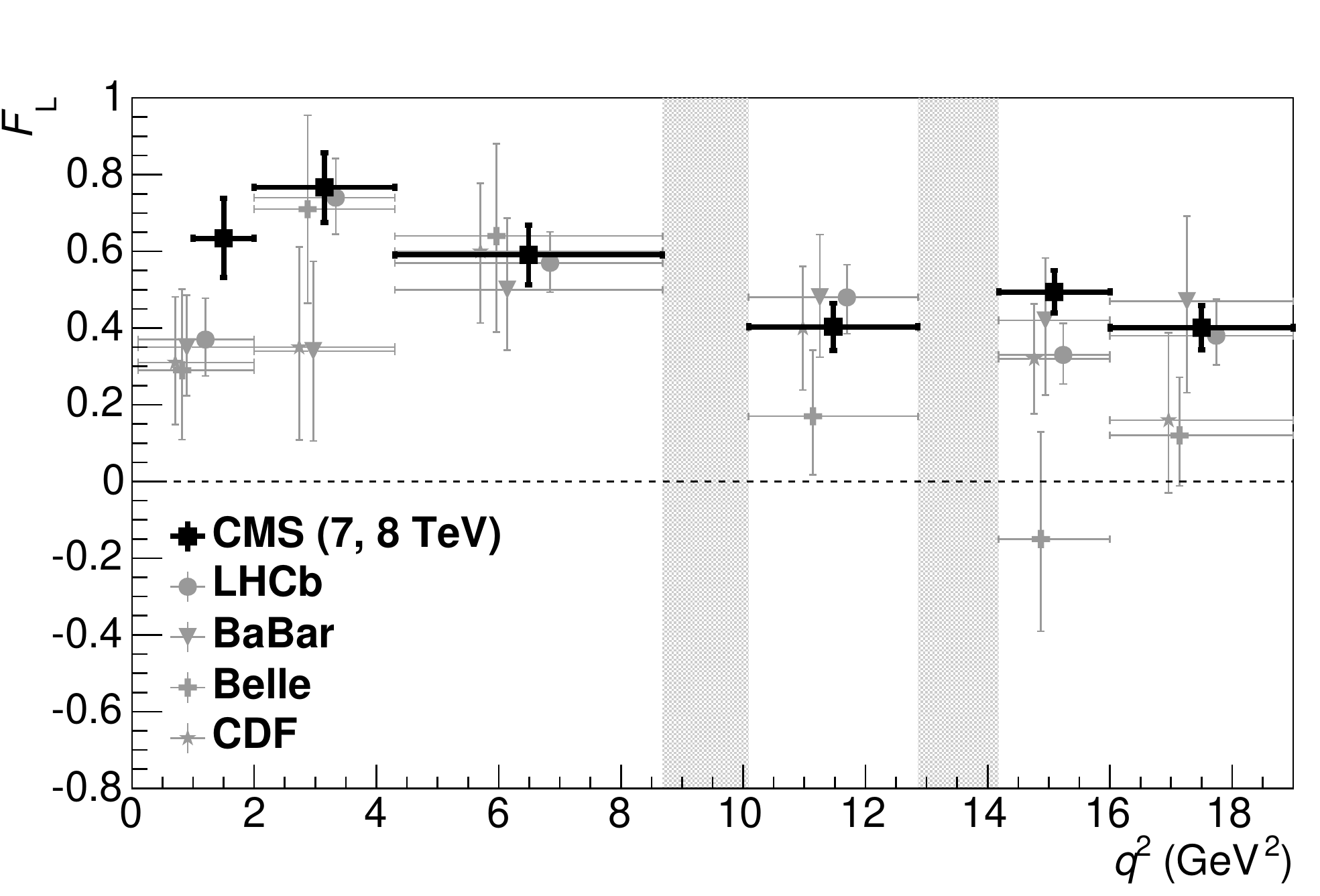}\hspace{5pt}
    \includegraphics[width=0.49\textwidth]{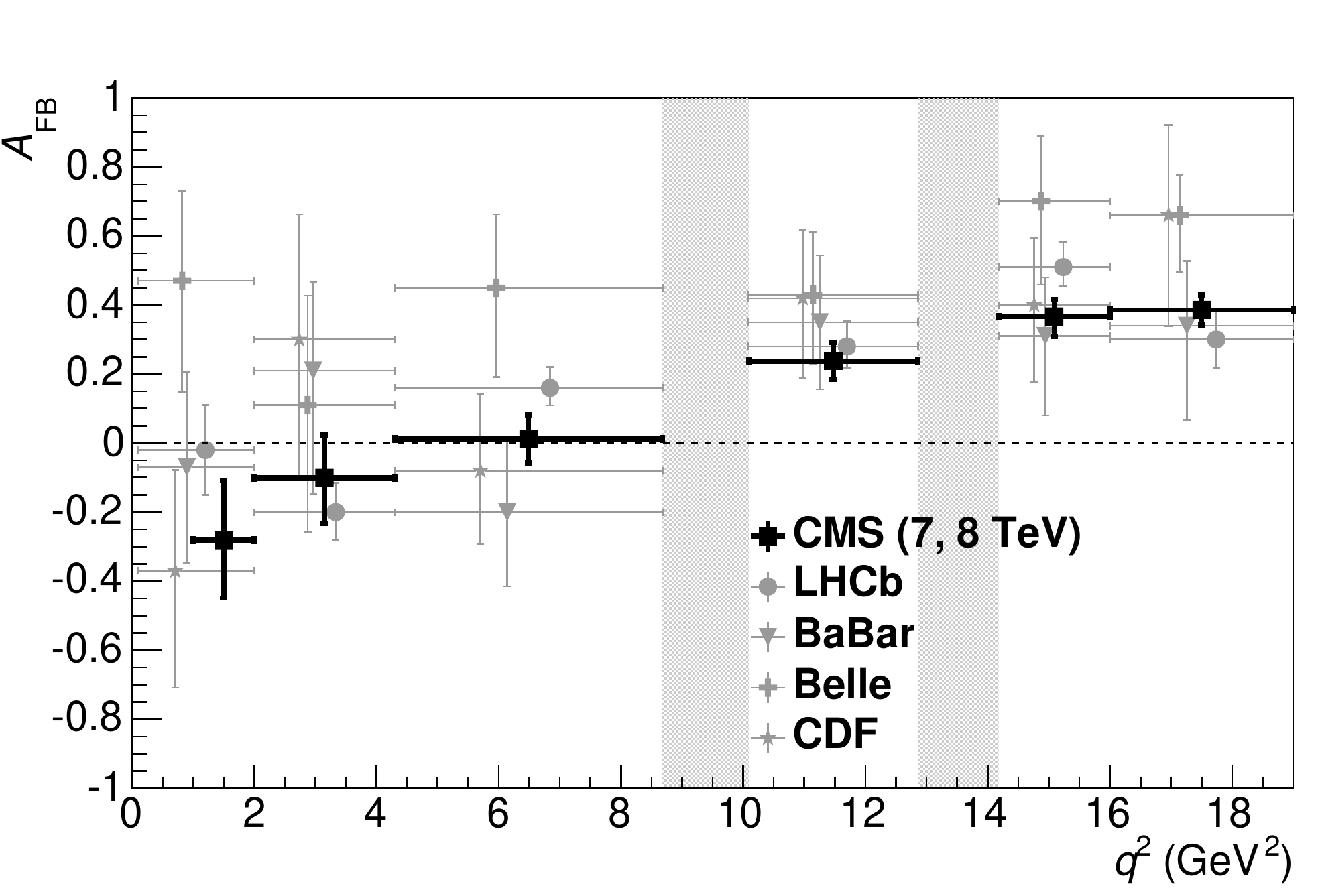}
    \includegraphics[width=0.49\textwidth]{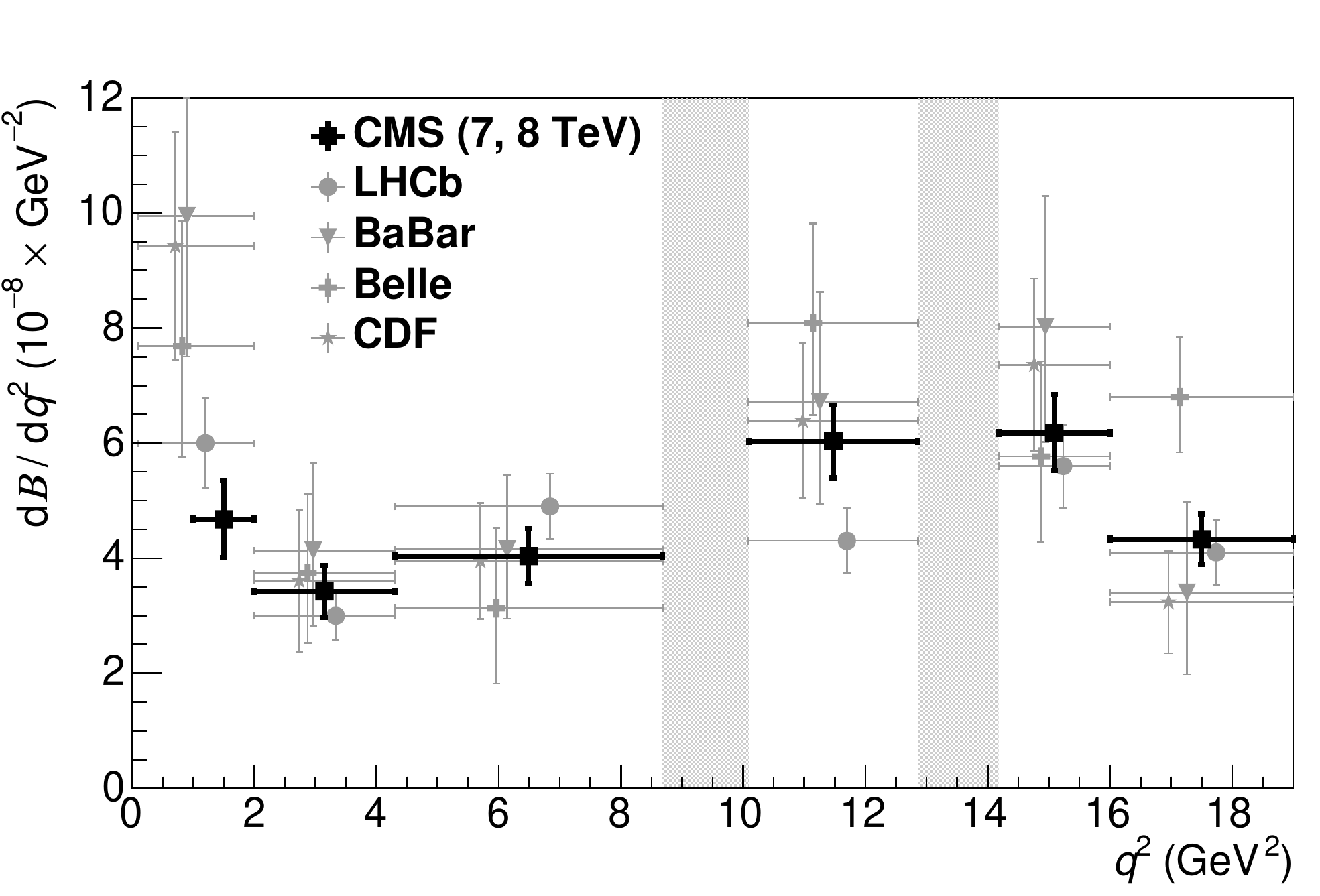}
    \caption{Measured values of $F_\mathrm{L}$, $A_\mathrm{FB}$, and $\rd{}\mathcal{B}/\rd{}q^2$ versus $q^2$
      for \BtoKstmumu from CMS (combination of the 7\TeV~\cite{Chatrchyan:2013cda} results and
      this analysis), Belle~\cite{Belle}, CDF~\cite{CDF,CDF_BR}, BaBar~\cite{BaBar_BR}, and
      LHCb~\cite{LHCb}.  The CMS and LHCb results are from \BtoKstmumu decays.  The remaining
      experiments add the corresponding $\PBp$ decay, and the BaBar and Belle experiments also
      include the dielectron mode.  The vertical bars give the total uncertainty.  The horizontal bars show the bin widths. The horizontal
      positions of the data points are staggered to improve legibility.  The vertical shaded regions
      correspond to the $\cPJgy$ and $\psi'$ resonances.}
    \label{fig:comp}
  \end{center}
\end{figure}

The results described are combined with previous CMS measurements, obtained from an independent data
sample collected at $\sqrt{s} = 7\TeV$~\cite{Chatrchyan:2013cda}.  The systematic uncertainties associated with the efficiency,
$K\pi$ mistagging, mass distribution, angular resolution, and the \BtoKstJpsi branching fraction are assumed to be fully
correlated between the two samples, with the remaining uncertainties assumed to be uncorrelated.  To combine the results from the 7\TeV and 8\TeV data, the
uncorrelated systematic uncertainties are combined in quadrature with the statistical uncertainties.
To account for the asymmetric uncertainties, the linear variance method from
Ref.~\cite{Barlow:2004wg} is used to average the 7\TeV and 8\TeV measurements, as well as to average
the two $q^2$ bins covering $4.30$ to $8.68\GeV^2$, which was a single bin in the 7\TeV
analysis.  After the combination, the correlated systematic uncertainties are added in
quadrature. The combined CMS measurements of $A_\mathrm{FB}$, $F_\mathrm{L}$, and the differential branching
fraction versus $q^2$ are compared to previous
measurements~\cite{Belle,CDF,CDF_BR,BaBar_BR,LHCb,Chatrchyan:2013cda} in Fig.~\ref{fig:comp}.  The
CMS measurements are consistent with the other results, with comparable or higher precision.  Table~\ref{tab:comp}
provides a comparison of the measured quantities in the low dimuon invariant mass region: $1 < q^2 <
6\GeV^2$, as well as the corresponding theoretical calculations.

\begin{table*}[!htbp]
  \centering
  \topcaption{\label{tab:comp}Measurements from CMS (the 7\TeV results~\cite{Chatrchyan:2013cda}, this work for 8\TeV, and
    the combination), LHCb~\cite{LHCb}, BaBar~\cite{BaBar_BR}, CDF~\cite{CDF,CDF_BR}, and
    Belle~\cite{Belle} of $F_\mathrm{L}$, $A_\mathrm{FB}$, and
    $\rd{}\mathcal{B}/\rd{}q^2$ in the region $1 < q^2 < 6\GeV^2$ for the decay
    \BtoKstmumu.  The CMS and LHCb results are from \BtoKstmumu decays.  The remaining experiments
    add the corresponding $\PBp$ decay, and the BaBar and Belle experiments also include the dielectron mode.
    The first uncertainty is statistical and the second is systematic.
    For the combined CMS results, only the total uncertainty is reported.  The two SM predictions are also
    given.}
\begin{tabular}{c|cccc}
Experiment & $F_\mathrm{L}$  & $A_\mathrm{FB}$ & $\rd{}\mathcal{B}/\rd{}q^2\;(10^{-8}\GeV^{-2})$ \\[1pt]
\hline
CMS (7\TeV)                & $0.68 \pm 0.10 \pm 0.02$ & $-0.07 \pm 0.12 \pm 0.01$             & $4.4\pm 0.6 \pm 0.4$ \\[1pt]
CMS (8\TeV, this analysis) & $0.73 \pm 0.05 \pm 0.04$ & $-0.16^{\:+\:0.10}_{\:-\:0.09} \pm 0.05$ & $ 3.6 \pm 0.3 \pm 0.2$ \\
CMS (7\TeV + 8\TeV)        & $0.72 \pm 0.06$          & $-0.12 \pm 0.08$                      & $ 3.8 \pm 0.4$ \\
LHCb     & $0.65^{\:+\:0.08}_{\:-\:0.07}\pm 0.03$ & $-0.17 \pm 0.06 \pm 0.01$ & $3.4\pm 0.3^{\:+\:0.4}_{\:-\:0.5}$ \\[1pt]
BaBar    & --- & --- & $4.1^{\:+\:1.1}_{\:-\:1.0}\pm 0.1$ \\[1pt]
CDF      & $0.69^{\:+\:0.19}_{\:-\:0.21} \pm 0.08$ & $0.29^{\:+\:0.20}_{\:-\:0.23} \pm 0.07$         & $3.2\pm 1.1 \pm 0.3$ \\[1pt]
Belle    & $0.67 \pm 0.23 \pm 0.05$  & $0.26^{\:+\:0.27}_{\:-\:0.32} \pm 0.07$    & $3.0^{\:+\:0.9}_{\:-\:0.8} \pm 0.2$ \\[1pt]
\hline
SM (LCSR)  & $0.79^{\:+\:0.09}_{\:-\:0.12}$ & $-0.02^{\:+\:0.03}_{\:-\:0.02}$ & $4.6^{\:+\:2.3}_{\:-\:1.7}$ \\[1pt]
SM (Lattice)  & $0.73^{\:+\:0.08}_{\:-\:0.10}$ & $-0.03^{\:+\:0.04}_{\:-\:0.03}$ & $3.8^{\:+\:1.2}_{\:-\:1.0}$ \\[1pt]
\end{tabular}
\end{table*}

\section{Summary}
\label{sec:End}

Using pp collision data recorded at $\sqrt{s}=8\TeV$ with the CMS detector at the LHC, corresponding to an integrated
luminosity of 20.5\fbinv, an angular analysis has been carried out on the decay \BtoKstmumu. The
data used for this analysis include 1430 signal decays.  For each bin of the dimuon invariant mass squared $(q^2)$,
unbinned maximum-likelihood fits were performed
to the distributions of the $\PKp\Pgpm\Pgmp\Pgmm$ invariant mass and two decay angles,
to obtain values of the forward-backward asymmetry of the muons, $A_\mathrm{FB}$, the fraction of
longitudinal polarization of the $\cPKstz$, $F_\mathrm{L}$, and the differential branching fraction,
$\rd{}\mathcal{B}/\rd{}q^2$. The results are among the most precise to date and are
consistent with standard model predictions and previous measurements.

\begin{acknowledgments}
We congratulate our colleagues in the CERN accelerator departments for the excellent performance of the LHC and thank the technical and administrative staffs at CERN and at other CMS institutes for their contributions to the success of the CMS effort. In addition, we gratefully acknowledge the computing centers and personnel of the Worldwide LHC Computing Grid for delivering so effectively the computing infrastructure essential to our analyses. Finally, we acknowledge the enduring support for the construction and operation of the LHC and the CMS detector provided by the following funding agencies: BMWFW and FWF (Austria); FNRS and FWO (Belgium); CNPq, CAPES, FAPERJ, and FAPESP (Brazil); MES (Bulgaria); CERN; CAS, MoST, and NSFC (China); COLCIENCIAS (Colombia); MSES and CSF (Croatia); RPF (Cyprus); MoER, ERC IUT and ERDF (Estonia); Academy of Finland, MEC, and HIP (Finland); CEA and CNRS/IN2P3 (France); BMBF, DFG, and HGF (Germany); GSRT (Greece); OTKA and NIH (Hungary); DAE and DST (India); IPM (Iran); SFI (Ireland); INFN (Italy); MSIP and NRF (Republic of Korea); LAS (Lithuania); MOE and UM (Malaysia); CINVESTAV, CONACYT, SEP, and UASLP-FAI (Mexico); MBIE (New Zealand); PAEC (Pakistan); MSHE and NSC (Poland); FCT (Portugal); JINR (Dubna); MON, RosAtom, RAS and RFBR (Russia); MESTD (Serbia); SEIDI and CPAN (Spain); Swiss Funding Agencies (Switzerland); MST (Taipei); ThEPCenter, IPST, STAR and NSTDA (Thailand); TUBITAK and TAEK (Turkey); NASU and SFFR (Ukraine); STFC (United Kingdom); DOE and NSF (USA).

Individuals have received support from the Marie-Curie program and the European Research Council and EPLANET (European Union); the Leventis Foundation; the A. P. Sloan Foundation; the Alexander von Humboldt Foundation; the Belgian Federal Science Policy Office; the Fonds pour la Formation \`a la Recherche dans l'Industrie et dans l'Agriculture (FRIA-Belgium); the Agentschap voor Innovatie door Wetenschap en Technologie (IWT-Belgium); the Ministry of Education, Youth and Sports (MEYS) of the Czech Republic; the Council of Science and Industrial Research, India; the HOMING PLUS program of the Foundation for Polish Science, cofinanced from European Union, Regional Development Fund; the Compagnia di San Paolo (Torino); the Consorzio per la Fisica (Trieste); MIUR project 20108T4XTM (Italy); the Thalis and Aristeia programs cofinanced by EU-ESF and the Greek NSRF; the National Priorities Research Program by Qatar National Research Fund; the Rachadapisek Sompot Fund for Postdoctoral Fellowship, Chulalongkorn University (Thailand); and the Welch Foundation.
\end{acknowledgments}
\bibliography{auto_generated}

\providecommand{\href}[2]{#2}\begingroup\raggedright\begin{thebibliography}{10}%
\makeatletter
\providecommand{\hrefCMSnoop }[0]{\@secondoftwo}%
\makeatother
\providecommand{\doi}{\texttt{doi:}\begingroup \urlstyle{tt}\Url}

\bibitem{Altmannshofer:2008dz}
W.~Altmannshofer\hrefCMSnoop {}{ {et~al.}, ``{Symmetries and asymmetries of $B
  \to K^{*} \mu^{+} \mu^{-}$ decays in the Standard Model and beyond}'',}
  \textit{ JHEP} \textbf{ 01} (2009) 019,
  \href{http://dx.doi.org/10.1088/1126-6708/2009/01/019}{\doi{10.1088/1126-6708/2009/01/019}},
\href{http://www.arXiv.org/abs/0811.1214}{\texttt{arXiv:0811.1214}}.

\bibitem{Melikhov:1998cd}
\hrefCMSnoop {}{D.~Melikhov, N.~Nikitin, and S.~Simula, ``{Probing right-handed
  currents in $B \to K^* \ell^+ \ell^-$ transitions}'',} \textit{ Phys. Lett.
  B} \textbf{ 442} (1998) 381,
  \href{http://dx.doi.org/10.1016/S0370-2693(98)01271-4}{\doi{10.1016/S0370-2693(98)01271-4}},
\href{http://www.arXiv.org/abs/hep-ph/9807464}{\texttt{arXiv:hep-ph/9807464}}.

\bibitem{Ali:1999mm}
\hrefCMSnoop {}{A.~Ali, P.~Ball, L.~T. Handoko, and G.~Hiller, ``{A comparative
  study of the decays $B \to (K, K^*) \ell^+ \ell^-$ in the standard model and
  supersymmetric theories}'',} \textit{ Phys. Rev. D} \textbf{ 61} (2000)
  074024,
  \href{http://dx.doi.org/10.1103/PhysRevD.61.074024}{\doi{10.1103/PhysRevD.61.074024}},
\href{http://www.arXiv.org/abs/hep-ph/9910221}{\texttt{arXiv:hep-ph/9910221}}.

\bibitem{Yan:2000dc}
\hrefCMSnoop {}{Q.-S. Yan, C.-S. Huang, W.~Liao, and S.-H. Zhu, ``{Exclusive
  semileptonic rare decays $B \to (K, K^*) \ell^+ \ell^-$ in supersymmetric
  theories}'',} \textit{ Phys. Rev. D} \textbf{ 62} (2000) 094023,
  \href{http://dx.doi.org/10.1103/PhysRevD.62.094023}{\doi{10.1103/PhysRevD.62.094023}},
\href{http://www.arXiv.org/abs/hep-ph/0004262}{\texttt{arXiv:hep-ph/0004262}}.

\bibitem{Buchalla:2000sk}
\hrefCMSnoop {}{G.~Buchalla, G.~Hiller, and G.~Isidori, ``{Phenomenology of
  nonstandard $Z$ couplings in exclusive semileptonic $b \to s$
  transitions}'',} \textit{ Phys. Rev. D} \textbf{ 63} (2000) 014015,
  \href{http://dx.doi.org/10.1103/PhysRevD.63.014015}{\doi{10.1103/PhysRevD.63.014015}},
\href{http://www.arXiv.org/abs/hep-ph/0006136}{\texttt{arXiv:hep-ph/0006136}}.

\bibitem{Feldmann:2002iw}
\hrefCMSnoop {}{T.~Feldmann and J.~Matias, ``{Forward-backward and isospin
  asymmetry for $B \to K^* \ell^+ \ell^-$ decay in the standard model and in
  supersymmetry}'',} \textit{ JHEP} \textbf{ 01} (2003) 074,
  \href{http://dx.doi.org/10.1088/1126-6708/2003/01/074}{\doi{10.1088/1126-6708/2003/01/074}},
\href{http://www.arXiv.org/abs/hep-ph/0212158}{\texttt{arXiv:hep-ph/0212158}}.

\bibitem{Hiller:2003js}
\hrefCMSnoop {}{G.~Hiller and F.~Kr{\"u}ger, ``{More model-independent analysis
  of $b \to s$ processes}'',} \textit{ Phys. Rev. D} \textbf{ 69} (2004)
  074020,
  \href{http://dx.doi.org/10.1103/PhysRevD.69.074020}{\doi{10.1103/PhysRevD.69.074020}},
\href{http://www.arXiv.org/abs/hep-ph/0310219}{\texttt{arXiv:hep-ph/0310219}}.

\bibitem{Kruger:2005ep}
\hrefCMSnoop {}{F.~Kr{\"u}ger and J.~Matias, ``{Probing new physics via the
  transverse amplitudes of $B^0 \to K^{*0}(\to K^-\pi^+)\ell^+\ell^-$ at large
  recoil}'',} \textit{ Phys. Rev. D} \textbf{ 71} (2005) 094009,
  \href{http://dx.doi.org/10.1103/PhysRevD.71.094009}{\doi{10.1103/PhysRevD.71.094009}},
\href{http://www.arXiv.org/abs/hep-ph/0502060}{\texttt{arXiv:hep-ph/0502060}}.

\bibitem{Hovhannisyan:2007pb}
\hrefCMSnoop {}{W.-S. Hou, A.~Hovhannisyan, and N.~Mahajan, ``{$B \to K^*
  \ell^+ \ell^-$ forward-backward asymmetry and new physics}'',} \textit{ Phys.
  Rev. D} \textbf{ 77} (2008) 014016,
  \href{http://dx.doi.org/10.1103/PhysRevD.77.014016}{\doi{10.1103/PhysRevD.77.014016}},
\href{http://www.arXiv.org/abs/hep-ph/0701046}{\texttt{arXiv:hep-ph/0701046}}.

\bibitem{Egede:2008uy}
U.~Egede\hrefCMSnoop {}{ {et~al.}, ``New observables in the decay mode
  $\overline{B}_d \to \overline{K}^{*0} \ell^+ \ell^-$'',} \textit{ JHEP}
  \textbf{ 11} (2008) 032,
  \href{http://dx.doi.org/10.1088/1126-6708/2008/11/032}{\doi{10.1088/1126-6708/2008/11/032}},
\href{http://www.arXiv.org/abs/0807.2589}{\texttt{arXiv:0807.2589}}.

\bibitem{Hurth:2008jc}
\hrefCMSnoop {}{T.~Hurth, G.~Isidori, J.~F. Kamenik, and F.~Mescia,
  ``{Constraints on new physics in MFV models: A model-independent analysis of
  $\Delta F = 1$ processes}'',} \textit{ Nucl. Phys. B} \textbf{ 808} (2009)
  326,
  \href{http://dx.doi.org/10.1016/j.nuclphysb.2008.09.040}{\doi{10.1016/j.nuclphysb.2008.09.040}},
\href{http://www.arXiv.org/abs/0807.5039}{\texttt{arXiv:0807.5039}}.

\bibitem{Alok:2009tz}
A.~K. Alok\hrefCMSnoop {}{ {et~al.}, ``{New-physics contributions to the
  forward-backward asymmetry in $B\to K^*\mu^+\mu^-$}'',} \textit{ JHEP}
  \textbf{ 02} (2010) 053,
  \href{http://dx.doi.org/10.1007/JHEP02(2010)053}{\doi{10.1007/JHEP02(2010)053}},
\href{http://www.arXiv.org/abs/0912.1382}{\texttt{arXiv:0912.1382}}.

\bibitem{Alok:2010zd}
A.~K. Alok\hrefCMSnoop {}{ {et~al.}, ``{New physics in $b \to s \mu^+ \mu^-$:
  CP-conserving observables}'',} \textit{ JHEP} \textbf{ 11} (2011) 121,
  \href{http://dx.doi.org/10.1007/JHEP11(2011)121}{\doi{10.1007/JHEP11(2011)121}},
\href{http://www.arXiv.org/abs/1008.2367}{\texttt{arXiv:1008.2367}}.

\bibitem{Chang:2010zy}
\hrefCMSnoop {}{Q.~Chang, X.-Q. Li, and Y.-D. Yang, ``{$B \to K^* \ell^+
  \ell^-$, $K \ell^+ \ell^-$ decays in a family non-universal $Z^{\prime}$
  model}'',} \textit{ JHEP} \textbf{ 04} (2010) 052,
  \href{http://dx.doi.org/10.1007/JHEP04(2010)052}{\doi{10.1007/JHEP04(2010)052}},
\href{http://www.arXiv.org/abs/1002.2758}{\texttt{arXiv:1002.2758}}.

\bibitem{DescotesGenon:2011yn}
\hrefCMSnoop {}{S.~Descotes-Genon, D.~Ghosh, J.~Matias, and M.~Ramon,
  ``{Exploring new physics in the C$_7$-C$_{7'}$ plane}'',} \textit{ JHEP}
  \textbf{ 06} (2011) 099,
  \href{http://dx.doi.org/10.1007/JHEP06(2011)099}{\doi{10.1007/JHEP06(2011)099}},
\href{http://www.arXiv.org/abs/1104.3342}{\texttt{arXiv:1104.3342}}.

\bibitem{Matias:2012xw}
\hrefCMSnoop {}{J.~Matias, F.~Mescia, M.~Ramon, and J.~Virto, ``{Complete
  anatomy of $\overline{B}_d \to \overline{K}^{* 0} (\to K \pi) \ell^+ \ell^-$
  and its angular distribution}'',} \textit{ JHEP} \textbf{ 04} (2012) 104,
  \href{http://dx.doi.org/10.1007/JHEP04(2012)104}{\doi{10.1007/JHEP04(2012)104}},
\href{http://www.arXiv.org/abs/1202.4266}{\texttt{arXiv:1202.4266}}.

\bibitem{DescotesGenon:2012zf}
\hrefCMSnoop {}{S.~Descotes-Genon, J.~Matias, M.~Ramon, and J.~Virto,
  ``{Implications from clean observables for the binned analysis of $B \to
  K^*\mu^+\mu^-$ at large recoil}'',} \textit{ JHEP} \textbf{ 01} (2013) 048,
  \href{http://dx.doi.org/10.1007/JHEP01(2013)048}{\doi{10.1007/JHEP01(2013)048}},
\href{http://www.arXiv.org/abs/1207.2753}{\texttt{arXiv:1207.2753}}.

\bibitem{Bobeth:2010wg}
\hrefCMSnoop {}{C.~Bobeth, G.~Hiller, and D.~van Dyk, ``The benefits of
  {$\overline{B} \to \overline{K}^* \ell^+ \ell^-$} decays at low recoil'',}
  \textit{ JHEP} \textbf{ 07} (2010) 098,
  \href{http://dx.doi.org/10.1007/JHEP07(2010)098}{\doi{10.1007/JHEP07(2010)098}},
\href{http://www.arXiv.org/abs/1006.5013}{\texttt{arXiv:1006.5013}}.

\bibitem{Bobeth:2011nj}
\hrefCMSnoop {}{C.~Bobeth, G.~Hiller, D.~van Dyk, and C.~Wacker, ``The decay
  $\overline{B} \to \overline{K} \ell^+ \ell^-$ at low hadronic recoil and
  model-independent $\Delta B = 1$ constraints'',} \textit{ JHEP} \textbf{ 01}
  (2012) 107,
  \href{http://dx.doi.org/10.1007/JHEP01(2012)107}{\doi{10.1007/JHEP01(2012)107}},
\href{http://www.arXiv.org/abs/1111.2558}{\texttt{arXiv:1111.2558}}.

\bibitem{Bobeth:2012vn}
\hrefCMSnoop {}{C.~Bobeth, G.~Hiller, and D.~van Dyk, ``General analysis of
  {$\overline{B} \to \overline{K}{}^{(*)} \ell^+ \ell^-$} decays at low
  recoil'',} \textit{ Phys. Rev. D} \textbf{ 87} (2012) 034016,
  \href{http://dx.doi.org/10.1103/PhysRevD.87.034016}{\doi{10.1103/PhysRevD.87.034016}},
\href{http://www.arXiv.org/abs/1212.2321}{\texttt{arXiv:1212.2321}}.

\bibitem{Ali:2006ew}
\hrefCMSnoop {}{A.~Ali, G.~Kramer, and G.~Zhu, ``{$B\to K^*\ell^+\ell^-$ decay
  in soft-collinear effective theory}'',} \textit{ Eur. Phys. J. C} \textbf{
  47} (2006) 625,
  \href{http://dx.doi.org/10.1140/epjc/s2006-02596-4}{\doi{10.1140/epjc/s2006-02596-4}},
\href{http://www.arXiv.org/abs/hep-ph/0601034}{\texttt{arXiv:hep-ph/0601034}}.

\bibitem{Altmannshofer:2011gn}
\hrefCMSnoop {}{W.~Altmannshofer, P.~Paradisi, and D.~M. Straub,
  ``{Model-independent constraints on new physics in $b \to s$ transitions}'',}
  \textit{ JHEP} \textbf{ 04} (2012) 008,
  \href{http://dx.doi.org/10.1007/JHEP04(2012)008}{\doi{10.1007/JHEP04(2012)008}},
\href{http://www.arXiv.org/abs/1111.1257}{\texttt{arXiv:1111.1257}}.

\bibitem{Jager:2012uw}
\hrefCMSnoop {}{S.~J{\"a}ger and J.~Martin~Camalich, ``{On $B \to V \ell \ell$
  at small dilepton invariant mass, power corrections, and new physics}'',}
  \textit{ JHEP} \textbf{ 05} (2013) 043,
  \href{http://dx.doi.org/10.1007/JHEP05(2013)043}{\doi{10.1007/JHEP05(2013)043}},
\href{http://www.arXiv.org/abs/1212.2263}{\texttt{arXiv:1212.2263}}.

\bibitem{Descotes-Genon:2013vna}
\hrefCMSnoop {}{S.~Descotes-Genon, T.~Hurth, J.~Matias, and J.~Virto,
  ``{Optimizing the basis of $B \to {K}^{*}\ell^+ \ell^-$ observables in the
  full kinematic range}'',} \textit{ JHEP} \textbf{ 05} (2013) 137,
  \href{http://dx.doi.org/10.1007/JHEP05(2013)137}{\doi{10.1007/JHEP05(2013)137}},
\href{http://www.arXiv.org/abs/1303.5794}{\texttt{arXiv:1303.5794}}.

\bibitem{BaBar}
\hrefCMSnoop {}{{BaBar} Collaboration, ``Angular distributions in the decay {$B
  \to K^* \ell^+ \ell^-$}'',} \textit{ Phys. Rev. D} \textbf{ 79} (2009)
  031102,
  \href{http://dx.doi.org/10.1103/PhysRevD.79.031102}{\doi{10.1103/PhysRevD.79.031102}},
\href{http://www.arXiv.org/abs/0804.4412}{\texttt{arXiv:0804.4412}}.

\bibitem{Belle}
\hrefCMSnoop {}{{Belle} Collaboration, ``Measurement of the differential
  branching fraction and forward-backward asymmetry for {$B \to K^{(*)} \ell^+
  \ell^-$}'',} \textit{ Phys. Rev. Lett.} \textbf{ 103} (2009) 171801,
  \href{http://dx.doi.org/10.1103/PhysRevLett.103.171801}{\doi{10.1103/PhysRevLett.103.171801}},
\href{http://www.arXiv.org/abs/0904.0770}{\texttt{arXiv:0904.0770}}.

\bibitem{CDF}
\hrefCMSnoop {}{{{CDF}} Collaboration, ``Measurements of the angular
  distributions in the decays {$B \to K^{(*)} \mu^+ \mu^-$} at {CDF}'',}
  \textit{ Phys. Rev. Lett.} \textbf{ 108} (2012) 081807,
  \href{http://dx.doi.org/10.1103/PhysRevLett.108.081807}{\doi{10.1103/PhysRevLett.108.081807}},
\href{http://www.arXiv.org/abs/1108.0695}{\texttt{arXiv:1108.0695}}.

\bibitem{LHCb}
\hrefCMSnoop {}{{{LHCb}} Collaboration, ``{Differential branching fraction and
  angular analysis of the decay $B^{0} \to K^{*0} \mu^{+}\mu^{-}$}'',} \textit{
  JHEP} \textbf{ 08} (2013) 131,
  \href{http://dx.doi.org/10.1007/JHEP08(2013)131}{\doi{10.1007/JHEP08(2013)131}},
\href{http://www.arXiv.org/abs/1304.6325}{\texttt{arXiv:1304.6325}}.

\bibitem{Chatrchyan:2013cda}
\hrefCMSnoop {}{{CMS Collaboration}, ``{Angular analysis and branching fraction
  measurement of the decay $B^0 \to K^{*0} \mu^+\mu^-$}'',} \textit{ Phys.
  Lett. B} \textbf{ 727} (2013) 77,
  \href{http://dx.doi.org/10.1016/j.physletb.2013.10.017}{\doi{10.1016/j.physletb.2013.10.017}},
\href{http://www.arXiv.org/abs/1308.3409}{\texttt{arXiv:1308.3409}}.

\bibitem{LUMI}
\href {http://cds.cern.ch/record/1598864}{{CMS Collaboration}, ``CMS luminosity
  based on pixel cluster counting - Summer 2013 update'',} CMS Physics Analysis
  Summary CMS-PAS-LUM-13-001, 2013.

\bibitem{CMS}
\hrefCMSnoop {}{{{CMS}} Collaboration, ``The {CMS} experiment at the {CERN}
  {LHC}'',} \textit{ JINST} \textbf{ 3} (2008) S08004,
\href{http://dx.doi.org/10.1088/1748-0221/3/08/S08004}{\doi{10.1088/1748-0221/3/08/S08004}}.

\bibitem{TRK-11-001}
\hrefCMSnoop {}{{CMS Collaboration}, ``{Description and performance of track
  and primary-vertex reconstruction with the CMS tracker}'',} \textit{ JINST}
  \textbf{ 9} (2014) P10009,
  \href{http://dx.doi.org/10.1088/1748-0221/9/10/P10009}{\doi{10.1088/1748-0221/9/10/P10009}},
\href{http://www.arXiv.org/abs/1405.6569}{\texttt{arXiv:1405.6569}}.

\bibitem{Chatrchyan:2012xi}
\hrefCMSnoop {}{{CMS Collaboration}, ``{Performance of CMS muon reconstruction
  in pp collision events at $\sqrt{s}=7$ TeV}'',} \textit{ JINST} \textbf{ 7}
  (2012) P10002,
  \href{http://dx.doi.org/10.1088/1748-0221/7/10/P10002}{\doi{10.1088/1748-0221/7/10/P10002}},
\href{http://www.arXiv.org/abs/1206.4071}{\texttt{arXiv:1206.4071}}.

\bibitem{PDG}
\hrefCMSnoop {}{{Particle Data Group}, K.~A. Olive {et~al.}, ``{The Review of
  Particle Physics}'',} \textit{ Chin. Phys. C} \textbf{ 38} (2014) 090001,
\href{http://dx.doi.org/10.1088/1674-1137/38/9/090001}{\doi{10.1088/1674-1137/38/9/090001}}.

\bibitem{Pythia}
\hrefCMSnoop {}{T.~Sj{\"o}strand, S.~Mrenna, and P.~Skands, ``{PYTHIA} 6.4
  physics and manual'',} \textit{ JHEP} \textbf{ 05} (2006) 026,
  \href{http://dx.doi.org/10.1088/1126-6708/2006/05/026}{\doi{10.1088/1126-6708/2006/05/026}},
\href{http://www.arXiv.org/abs/hep-ph/0603175}{\texttt{arXiv:hep-ph/0603175}}.

\bibitem{EvtGen}
\hrefCMSnoop {}{D.~J. Lange, ``The {EvtGen} particle decay simulation
  package'',} \textit{ Nucl. Instrum. Meth. A} \textbf{ 462} (2001) 152,
\href{http://dx.doi.org/10.1016/S0168-9002(01)00089-4}{\doi{10.1016/S0168-9002(01)00089-4}}.

\bibitem{Geant4}
\hrefCMSnoop {}{{GEANT4} Collaboration, ``{GEANT4}---a simulation toolkit'',}
  \textit{ Nucl. Instrum. Meth. A} \textbf{ 506} (2003) 250,
\href{http://dx.doi.org/10.1016/S0168-9002(03)01368-8}{\doi{10.1016/S0168-9002(03)01368-8}}.

\bibitem{Becirevic:2012dp}
\hrefCMSnoop {}{D.~Be\v{c}irevi\'{c} and A.~Tayduganov, ``{Impact of $B\to
  K^{*0} \ell^+\ell^-$ on the New Physics search in $B\to K^* \ell^+\ell^-$
  decay}'',} \textit{ Nucl. Phys. B} \textbf{ 868} (2013) 368,
  \href{http://dx.doi.org/10.1016/j.nuclphysb.2012.11.016}{\doi{10.1016/j.nuclphysb.2012.11.016}},
\href{http://www.arXiv.org/abs/1207.4004}{\texttt{arXiv:1207.4004}}.

\bibitem{Matias:2012qz}
\hrefCMSnoop {}{J.~Matias, ``{On the S-wave pollution of $B \to K^*
  \ell^+\ell^-$ observables}'',} \textit{ Phys. Rev. D} \textbf{ 86} (2012)
  094024,
  \href{http://dx.doi.org/10.1103/PhysRevD.86.094024}{\doi{10.1103/PhysRevD.86.094024}},
\href{http://www.arXiv.org/abs/1209.1525}{\texttt{arXiv:1209.1525}}.

\bibitem{Blake:Swave}
\hrefCMSnoop {}{T.~Blake, U.~Egede, and A.~Shires, ``The effect of {$S$}-wave
  interference on the {$B^0 \rightarrow K^{*0} \ell^+ \ell^-$} angular
  observables'',} \textit{ JHEP} \textbf{ 03} (2013) 027,
  \href{http://dx.doi.org/10.1007/JHEP03(2013)027}{\doi{10.1007/JHEP03(2013)027}},
  \href{http://www.arXiv.org/abs/1210.5279}{\texttt{arXiv:1210.5279}}.

\bibitem{Minuit}
\hrefCMSnoop {}{F.~James and M.~Roos, ``{Minuit---a system for function
  minimization and analysis of the parameter errors and correlations}'',}
  \textit{ Comput. Phys. Commun.} \textbf{ 10} (1975) 343,
\href{http://dx.doi.org/10.1016/0010-4655(75)90039-9}{\doi{10.1016/0010-4655(75)90039-9}}.

\bibitem{Aaij:2012dda}
\hrefCMSnoop {}{{{LHCb}} Collaboration, ``{Measurement of relative branching
  fractions of B decays to $\psi(2S)$ and $J/\psi$ mesons}'',} \textit{ Eur.
  Phys. J. C} \textbf{ 72} (2012) 2118,
  \href{http://dx.doi.org/10.1140/epjc/s10052-012-2118-7}{\doi{10.1140/epjc/s10052-012-2118-7}},
\href{http://www.arXiv.org/abs/1205.0918}{\texttt{arXiv:1205.0918}}.

\bibitem{Beneke:2001at}
\hrefCMSnoop {}{M.~Beneke, T.~Feldmann, and D.~Seidel, ``Systematic approach to
  exclusive {$B \to V \ell^+ \ell^-, V \gamma$} decays'',} \textit{ Nucl. Phys.
  B} \textbf{ 612} (2001) 25,
  \href{http://dx.doi.org/10.1016/S0550-3213(01)00366-2}{\doi{10.1016/S0550-3213(01)00366-2}},
\href{http://www.arXiv.org/abs/hep-ph/0106067}{\texttt{arXiv:hep-ph/0106067}}.

\bibitem{Grinstein:2004vb}
\hrefCMSnoop {}{B.~Grinstein and D.~Pirjol, ``Exclusive rare {$B \to K^* \ell^+
  \ell^-$} decays at low recoil: controlling the long-distance effects'',}
  \textit{ Phys. Rev. D} \textbf{ 70} (2004) 114005,
  \href{http://dx.doi.org/10.1103/PhysRevD.70.114005}{\doi{10.1103/PhysRevD.70.114005}},
\href{http://www.arXiv.org/abs/hep-ph/0404250}{\texttt{arXiv:hep-ph/0404250}}.

\bibitem{Beylich:2011aq}
\hrefCMSnoop {}{M.~Beylich, G.~Buchalla, and T.~Feldmann, ``{Theory of $B \to
  K^{(*)}\ell^+\ell^-$ decays at high $q^2$: OPE and quark-hadron duality}'',}
  \textit{ Eur. Phys. J. C} \textbf{ 71} (2011) 1635,
  \href{http://dx.doi.org/10.1140/epjc/s10052-011-1635-0}{\doi{10.1140/epjc/s10052-011-1635-0}},
\href{http://www.arXiv.org/abs/1101.5118}{\texttt{arXiv:1101.5118}}.

\bibitem{Grinstein:2002cz}
\hrefCMSnoop {}{B.~Grinstein and D.~Pirjol, ``{Symmetry-breaking corrections to
  heavy meson form-factor relations}'',} \textit{ Phys. Lett. B} \textbf{ 533}
  (2002) 8,
  \href{http://dx.doi.org/10.1016/S0370-2693(02)01601-5}{\doi{10.1016/S0370-2693(02)01601-5}},
\href{http://www.arXiv.org/abs/hep-ph/0201298}{\texttt{arXiv:hep-ph/0201298}}.

\bibitem{Khodjamirian:2010}
\hrefCMSnoop {}{A.~Khodjamirian, T.~Mannel, A.~A. Pivovarov, and Y.-M. Wang,
  ``Charm-loop effect in ${B \rightarrow K^{(*)} \ell^+ \ell^-}$ and $ B
  \rightarrow {K}^*~\gamma$'',} \textit{ JHEP} \textbf{ 09} (2010) 089,
  \href{http://dx.doi.org/10.1007/JHEP09(2010)089}{\doi{10.1007/JHEP09(2010)089}},
  \href{http://www.arXiv.org/abs/1006.4945}{\texttt{arXiv:1006.4945}}.

\bibitem{Khodjamirian:2012rm}
\hrefCMSnoop {}{A.~Khodjamirian, T.~Mannel, and Y.-M. Wang, ``{$B \to K
  \ell^{+}\ell^{-}$ decay at large hadronic recoil}'',} \textit{ JHEP} \textbf{
  02} (2013) 010,
  \href{http://dx.doi.org/10.1007/JHEP02(2013)010}{\doi{10.1007/JHEP02(2013)010}},
\href{http://www.arXiv.org/abs/1211.0234}{\texttt{arXiv:1211.0234}}.

\bibitem{Horgan:2013hoa}
\hrefCMSnoop {}{R.~R. Horgan, Z.~Liu, S.~Meinel, and M.~Wingate, ``{Lattice QCD
  calculation of form factors describing the rare decays $B \to K^* \ell^+
  \ell^-$ and $B_s \to \phi \ell^+ \ell^-$}'',} \textit{ Phys. Rev. D} \textbf{
  89} (2014) 094501,
  \href{http://dx.doi.org/10.1103/PhysRevD.89.094501}{\doi{10.1103/PhysRevD.89.094501}},
\href{http://www.arXiv.org/abs/1310.3722}{\texttt{arXiv:1310.3722}}.

\bibitem{Barlow:2004wg}
\hrefCMSnoop {}{R.~Barlow, ``Asymmetric statistical errors'',} in \textit{
  Statistical problems in particle physics, astrophysics and cosmology: PHYSTAT
  05, Oxford, UK, September 12-15, 2005}, L.~Lyons and M.~Karagoz, eds., p.~56.
\newblock Imperial College Press, 2006.
\newblock
\href{http://www.arXiv.org/abs/physics/0406120}{\texttt{arXiv:physics/0406120}}.
\newblock

\bibitem{CDF_BR}
\hrefCMSnoop {}{{{CDF}} Collaboration, ``{Measurement of the forward-backward
  asymmetry in the $B \to K^{(*)} \mu^+ \mu^-$ decay and first observation of
  the $B^0_s \to \phi \mu^+ \mu^-$ decay}'',} \textit{ Phys. Rev. Lett.}
  \textbf{ 106} (2011) 161801,
  \href{http://dx.doi.org/10.1103/PhysRevLett.106.161801}{\doi{10.1103/PhysRevLett.106.161801}},
\href{http://www.arXiv.org/abs/1101.1028}{\texttt{arXiv:1101.1028}}.

\bibitem{BaBar_BR}
\hrefCMSnoop {}{{BaBar} Collaboration, ``{Measurement of branching fractions
  and rate asymmetries in the rare decays $B \to K^{(*)} \ell^+ \ell^-$}'',}
  \textit{ Phys. Rev. D} \textbf{ 86} (2012) 032012,
  \href{http://dx.doi.org/10.1103/PhysRevD.86.032012}{\doi{10.1103/PhysRevD.86.032012}},
\href{http://www.arXiv.org/abs/1204.3933}{\texttt{arXiv:1204.3933}}.

\end{thebibliography}\endgroup
\cleardoublepage \appendix\section{The CMS Collaboration \label{app:collab}}\begin{sloppypar}\hyphenpenalty=5000\widowpenalty=500\clubpenalty=5000\textbf{Yerevan Physics Institute,  Yerevan,  Armenia}\\*[0pt]
V.~Khachatryan, A.M.~Sirunyan, A.~Tumasyan
\vskip\cmsinstskip
\textbf{Institut f\"{u}r Hochenergiephysik der OeAW,  Wien,  Austria}\\*[0pt]
W.~Adam, E.~Asilar, T.~Bergauer, J.~Brandstetter, E.~Brondolin, M.~Dragicevic, J.~Er\"{o}, M.~Flechl, M.~Friedl, R.~Fr\"{u}hwirth\cmsAuthorMark{1}, V.M.~Ghete, C.~Hartl, N.~H\"{o}rmann, J.~Hrubec, M.~Jeitler\cmsAuthorMark{1}, V.~Kn\"{u}nz, A.~K\"{o}nig, M.~Krammer\cmsAuthorMark{1}, I.~Kr\"{a}tschmer, D.~Liko, T.~Matsushita, I.~Mikulec, D.~Rabady\cmsAuthorMark{2}, B.~Rahbaran, H.~Rohringer, J.~Schieck\cmsAuthorMark{1}, R.~Sch\"{o}fbeck, J.~Strauss, W.~Treberer-Treberspurg, W.~Waltenberger, C.-E.~Wulz\cmsAuthorMark{1}
% ...
\vskip\cmsinstskip
% ...
1:~~Also at Vienna University of Technology, Vienna, Austria\\
2:~~Also at CERN, European Organization for Nuclear Research, Geneva, Switzerland\\
% ...

\end{sloppypar}
\end{document}